\documentclass[aps,twocolumn,superscriptaddress,pra,floatfix,showpacs]{revtex4}
\usepackage[usenames]{color}
\usepackage{amsmath}
\usepackage{amssymb}
\usepackage{mathrsfs}
\usepackage{euscript}
\usepackage{graphicx}
\usepackage{graphics}
\begin{document}
\title{Theoretical study of the quenching of NH ($^1\Delta$) molecules via collisions with Rb atoms: preliminary results}
\author{Daniel J. Haxton}
\email{dhaxton@jila.colorado.edu}
\affiliation{Department of Physics and JILA, University of Colorado, Boulder Colorado 80309}

\author{Steven A. Wrathmall}
\email{steven.wrathmall@jila.colodado.edu}
\affiliation{Department of Physics and JILA, University of Colorado, Boulder Colorado 80309}

\author{H. J. Lewandowski}
\email{lewandoh@jilau1.colorado.edu}
\affiliation{Department of Physics and JILA, University of Colorado, Boulder Colorado 80309}

\author{Chris H. Greene}
\email{chris.greene@colorado.edu}
\affiliation{Department of Physics and JILA, University of Colorado, Boulder Colorado 80309}

\begin{abstract}

We examine the quenching reaction Rb ($^2$S) + NH ($^1\Delta$) $\rightarrow$ Rb ($^2$P$_{1/2}$) + NH ($X \ ^3\Sigma^-$).  This reaction may be
utilized to produce ground state NH molecules for studies of ultracold physics or for other purposes, and is interesting in that it
involves initial and final states that are nearly degenerate.  This near degeneracy is expected to lead to a large reaction rate.  We examine this
system using \textit{ab initio} quantum chemistry calculations and scattering calculations, which include spin-orbit effects, and find that the
reaction rate is large and, in fact, approaches the quantum mechanical unitarity limit.  We discuss the prospects for an experimental examination of this system.

\end{abstract}

\maketitle

\section{Introduction}

Chemical reactions involving free-radical molecules are important in many dense gas systems including combustion gases\cite{heather1} and interstellar
clouds\cite{heather2,heather4}.  The fundamental imidogen free radical(NH) has been detected \cite{heather5} as an intermediate in the combustion of CH$_4$ with N$_2$O. NH has a
ground state triplet, $X ^3\Sigma^-$, and a long lived metastable state\cite{heather6}, the singlet delta, labeled $a\Delta^1$ or $^1\Delta$. This state is doubly forbidden to decay via electric
dipole radiation ($^1\Delta \rightarrow$ $X ^3\Sigma^-$).
 Therefore, it is a good system in which to study properties of
excited molecular states\cite{heather7}.

Currently, there is considerable interest in collisions between diatomic molecules, such as NH, and alkali atoms. This interest stems from the
relevance of such systems to ultracold physics and chemistry~\cite{chemultra, coldkind,makingit, BEmol, labframe, nearzero, FH2} and the increased
level of control possible through the use of cooling and trapping techniques. The advent~\cite{BEC} of the Bose-Einsten condensate (BEC) led to
the production of mixed BECs~\cite{overlapping} and engendered work on production and trapping of cold molecules~\cite{cah, BECdipole, KRb,  opticalmol}. There has
been considerable recent experimental work on creating cold dipolar molecular samples for collision studies. For example, cold OH~\cite{trapping_oh}
and NH molecules in either the ground ($^3\Sigma^-$) or the $^1\Delta$ electronic state~\cite{magNH2,bufferNH, magNH, trapping_nh,cotrapping_n_nh}
have been produced. In addition, there are a growing number of theoretical investigations on interactions between diatomic molecules and alkali
atoms. A broad survey of interactions of alkali and alkaline earth metals with NH was published in Ref.~\cite{hutson_hope}.  
Other theoretical work on the Rb-NH system involved examinations of the relevant potential energy surfaces~\cite{hutson_rbnh,
gianturco_ultralow, gianturco_accounts}.  Explorations have also been conducted for related systems such as Rb-OH~\cite{hutson_rboh_prospects,
hutson_rboh}, He-NH~\cite{HeNHab, HeNH, hutson_henh} Rb-NH$_3$~\cite{hutson_nh3, hutson_nh3_xxx}, and NH-NH~\cite{NHNH}.

Our interest aims toward a broad treatment of the Rb-NH system, including up to the first excited electronic states of both the atom and molecule.  
We have constructed a preliminary treatment and present some initial results here.  The goal is to accurately calculate the dynamics on the higher excited electronic states.  In particular, we concentrate on the states $^3\Sigma^-$ and $^1\Delta$ of NH and the states $^2S$, $^2P_{1/2}$, and $^2P_{3/2}$ of Rb.   As it happens, the excitation
energies from the $^2S$ state to the $^2P$ states are nearly equal to the excitation energy of NH, such that the channels are unusually close in energy;
in fact, the excited NH channel lies between the two spin-orbit components of the Rb $^2P$ state.  Specifically, the excitation energies are
12687.8cm$^{-1}$ on the NH~\cite{data2}, and 12578.96 and 12816.55cm$^{-1}$ for Rb
$^2$P$_{1/2}$ and $^2$P$_{3/2}$, respectively~\cite{data1}.

This situation makes the quenching of the excited NH ($^1\Delta$) state
by Rb, using the reaction
\begin{equation}
\mathrm{Rb} (^2\mathrm{S}) + \mathrm{NH} ( ^1\Delta ) \quad \rightarrow \quad  \mathrm{Rb} (^2\mathrm{P}) + \mathrm{NH} (X ^3\Sigma^-) \ ,
\end{equation}
an interesting one to study for several reasons.  First, it opens 
up the possibility that this quenching reaction could find some use in ultracold
physics, as it would represent an electronically-inelastic reaction at cold initial and final collision energies. This is a chemically-reactive process, in the sense that the chemical nature of the products is dramatically different from that of the reactants. Second, it could prove useful as a mechanism for producing ground state NH from the exited state.

This quenching reaction can be precisely studied 
using techniques to cool and trap atomic and molecular 
samples. The methods of atom cooling and trapping have 
been refined over the last 15 years such
that one can routinely produce dense samples (up to 10$^{13}$ cm$^{-3}$) with temperatures on the order of 1--100 $\mu$K.  More recently, with the
development of Stark deceleration of polar molecules\cite{heather8}, trapped molecular samples can be produced with temperatures on the order of 1 --100 mK.

An experimental realization will proceed as follows. The atoms will be cooled in a magneto-optical trap and transferred to quadrupole magnetic trap in
one region of the vacuum system. During this process, a pulsed beam of NH molecules will be slowed using a Stark 
decelerator and trapped using electrostatic fields~\cite{heather9}.
Because of the weak interaction between the atom and the electric field, the two species can be controlled independently. This control will 
allow the atoms to be moved
to overlap the molecular sample, setting the initial time for the interaction. There will be several methods of detection employed in order to gain a full
understanding of the interaction. A novel signature of this reaction is the photon produced from the decay of the excited atom
($^2$P$_{1/2}$ $\rightarrow$ $^2$S$_{1/2}$). This dipole-allowed transition will happen very rapidly after the collisional excitation of Rb. These photons can be
detected efficiently with almost no background using a spectrally filtered photomultiplier tube. 
In addition, both the number and temperature of the trapped atom sample can be measured with absorption imaging;
the number and temperature of the molecular sample, with REMPI.
Through this procedure, it should be possible to measure the reaction rate.

\section{Quantum chemistry calculations}

We carry out quantum chemistry calculations using the COLUMBUS package~\cite{col1,col2,col3} for
calculations on the electronic states of the Rb-NH system.  We use an
effective core potential~\cite{pseudo1,pseudo2} that accounts for the 36 core electrons
of Rb, and thus treats the Rb atom as a one-electron system.  We therefore
perform calculations on nine-electron states.  We adjust our effective core potential to 
reproduce the correct channel energies of the combined Rb-NH system at infinite separation.

For the moment we have constructed a set of two-dimensional potential
energy surfaces, fixing the NH bond length at 1.925$a_0$.
We refer to the configuration of the Rb-NH molecule using Jacobi
coordinates.  These coordinates are defined as $r$, the NH bond distance,
1.925$a_0$; $R$, the distance between the NH center of mass
and the Rb; and $\gamma$, the angle between the two corresponding vectors,
such that $\gamma$=0 denotes a linear Rb-H-N configuration.

\begin{table}
\caption{Electronic microstates considered in our study: quantum numbers at linear geometry.  
Each is a member of a different Kramers doublet.
The quantum numbers also label the members of our diabatic basis for all geometries and the electronic channels of the
scattering calculation.  The quantum numbers
$\vert\Omega\vert$ and $\Sigma$ are the absolute values of the 
projection of the total and spin angular momentum on the NH axis. \label{so_diatable}}
\
\begin{ruledtabular}
\begin{tabular}{rlrl}
\hline
\multicolumn{2}{l}{\textbf{Doublets}} & 
1) & $^2\Delta$ ($^1\Delta$ $\times$ $^2S$) $ \vert\Omega\vert=\frac{3}{2} $  \\
2) & $^2\Delta$ ($^1\Delta$ $\times$ $^2S$) $ \vert\Omega\vert=\frac{5}{2} $  &
3) & $^2\Sigma$ ($^3\Sigma^-$ $\times$ $^2S$)  \\
4) & $^2\Pi$ anion $ \vert\Omega\vert=\frac{1}{2} $  &
5) & $^2\Pi$ anion $ \vert\Omega\vert=\frac{3}{2} $  \\
\hline
\multicolumn{2}{l}{\textbf{NH} $\mathbf{^3\Sigma^-}$ $\mathbf\times$ \textbf{Rb} $\mathbf{^2P}$, $\mathbf{\vert\Omega\vert=\frac{1}{2}}$}  &
6) & $\Sigma^{NH}=0$ \ Rb $^2P_{\frac{3}{2}}^{(\frac{1}{2})}$   \\
7) & $\Sigma^{NH}=1$ \  Rb $^2P_{\frac{3}{2}}^{(\frac{1}{2})}$   &
8) & $\Sigma^{NH}=0$ \  Rb $^2P_{\frac{1}{2}}^{(\frac{1}{2})}$   \\
9) & $\Sigma^{NH}=1$  \ Rb $^2P_{\frac{3}{2}}^{(\frac{3}{2})}$   &
10) & $\Sigma^{NH}=1$ \  Rb $^2P_{\frac{1}{2}}^{(\frac{1}{2})}$   \\
\hline
\multicolumn{2}{l}{ \textbf{NH} $\mathbf{^3\Sigma^-}$ $\mathbf\times$ \textbf{Rb} $\mathbf{^2P}$, $\mathbf{\vert\Omega\vert=\frac{3}{2}}$} &
11)& $\Sigma^{NH}=0$ \  Rb $^2P_{\frac{3}{2}}^{(\frac{3}{2})}$   \\
12)& $\Sigma^{NH}=1$ \  Rb $^2P_{\frac{1}{2}}^{(\frac{1}{2})}$   &
13)& $\Sigma^{NH}=1$ \  Rb $^2P_{\frac{3}{2}}^{(\frac{1}{2})}$   \\
\hline
\multicolumn{2}{l}{ \textbf{NH} $\mathbf{^3\Sigma^-}$ $\mathbf\times$ \textbf{Rb} $\mathbf{^2P}$, $\mathbf{\vert\Omega\vert=\frac{5}{2}}$} &
14) & $\Sigma^{NH}=1$ \  Rb $^2P_{\frac{3}{2}}^{(\frac{3}{2})}$   \\
\hline
\multicolumn{2}{l}{ $\mathbf{^4\Sigma}$ \textbf{(NH} $\mathbf{^3\Sigma^-}$ $\mathbf\times$ \textbf{Rb} $\mathbf{^2S}$) }\\
15) &  $\Sigma=\vert\Omega\vert=\frac{1}{2}$  &
16) &  $\Sigma=\vert\Omega\vert=\frac{3}{2}$  \\
\end{tabular}
\end{ruledtabular}
\end{table}

Owing to our use of a pseudopotential, in addition to the large number of electronic
states treated by a relatively modest configuration-interaction (CI) calculation, the surfaces we construct are not expected to be of ``spectroscopic accuracy'' but instead probably have errors on the order of tens of meV.  They are expected to be
sufficiently accurate for the qualitative study presented here, and in any case, the omission of motion in the Jacobi $r_{NH}$ degree of freedom surely compromises the calculated dynamics more than does any error in the potential energy surfaces.

In all, and without considering spin-orbit coupling, there are eight
doublet states and four quartet states, for a total of 32 microstates.  This odd
electron system enjoys the Kramers degeneracy and thus the electronic Hilbert space
is block-diagonal in two 16$\times$16 blocks.  For total angular momentum $J=0$
we need only consider one of these blocks.
The asymptotes of the adiabatic states are listed in Table~\ref{so_diatable}.

At linear geometry, the electronic states are described by their projections of angular
momentum on the molecular axis.
We refer to the electronic states by the standard $\Lambda$, $\Omega$, and $\Sigma$ quantum numbers.
$\Lambda$ is the absolute value of the 
projection of electronic angular momentum on the molecular axis, the eigenvalue
of the operator $\widehat{l_z}$, such that $\Lambda$=2 denotes a $\Delta$ state, etc.  
$\vert\Omega\vert$ is the absolute value of the projection of the total angular momentum on the
molecular axis, the eigenvalue of the operator $\widehat{j_z}$ = $\widehat{l_z}$ + $\widehat{s_z}$.
The absolute value of the eigenvalue of $\widehat{s_z}$ is denoted $\Sigma$.
The symbol $\Sigma$ is also used to denote a state with $\Lambda$=0, but we will avoid
ambiguity by always denoting the multiplicity of a given electronic state, as in $^2\Sigma$.

The Born-Oppenheimer electronic states are first calculated without spin orbit coupling.
The next step is a diabatization~\cite{diareview} of these potentials, in which we
construct diabatic states labeled primarily by integer values of $\Lambda$, by diagonalizing 
the electronic angular momentum $\widehat{l_z}$ projected onto the NH axis.  $\pm\Lambda$ is thus not only
the eigenvalue of the projection of angular momentum at linear geometry for a given diabatic electronic
state, but also its approximate expectation value at other geometries.
We have a pair of $^2\Delta$
diabatic states, two pairs of $^2\Pi$ states, and two $^2\Sigma$ states.  The doublet $\Pi$
states are distinguished as anionic in character (NH$^-$ $\times$ Rb$^+$ ) or not (NH $^3\Sigma^-$ $\times$ Rb $^2P$).  The $\Sigma$ states are
essentially NH $^3\Sigma^-$ times Rb in either the ground $^2S$ state or the $^2P_\sigma$ state.  We also have two $^4\Sigma$ and one $^4\Pi$ state.

A diabatization is useful to us for
two reasons.  First, it allows for the convenient inclusion of nonadiabatic effects among the adiabatic surfaces,
accounting for conical intersections, for instance. Rovibronic effects such as the Renner-Teller
effect can also be accounted for by such a diabatization, though at this stage we do not include the
mixing of electronic and rotational angular momentum.

Second, it appears to allow us to add the
spin-orbit terms ``by hand,'' and therefore enables us to perform a larger electronic structure calculation, though we employ a smaller calculation using the spin-orbit CI capability of COLUMBUS to verify the accuracy of this procedure.  The final steps in constructing the surfaces
are then the addition of spin-orbit terms to the diabatic electronic Hamiltonian
and, last, transformation to a representation that accounts for the Kramers
degeneracy.

\subsection{Details}

We have found it prohibitive to use a basis set large enough to get a sufficiently accurate NH
excitation energy.
Therefore, in order to reproduce the physical energetics,
we artificially modify
the effective Rb core potential.
We begin with the SDF pseudopotential developed in Refs.~\cite{pseudo1,pseudo2}.
The pseudopotential for the $S$ and $P$ waves is then modified by changing 
their functional forms into:
\begin{equation}
\begin{split}
& s: \\
& 45.272 \exp(-1.012 r^2 ) \rightarrow  45.272 \exp(-1.14973057 r^2 ) \\
& p: \\
& 2.83 \exp(-0.3036 r^2 ) \rightarrow  2.83 \exp(-0.267230608 r^2 )
\end{split}
\end{equation}

An uncontracted SDF basis set is adopted for the Rb; while for N and H we use
the aug-cc-pvtz basis set of Dunning~\cite{dunning1}.  The excitation energies obtained
are approximately 1298(5) cm$^{-1}$ on both the Rb (without spin-orbit interaction) and NH fragments.

The electronic configuration of the NH states of interest is
$1s^2 2s^2 2p_\sigma^2 2p_\pi^2$.  In addition to these orbitals we have
the four orbitals on rubidium ($s$ and $p$), for a total of nine orbitals
in the valence space.

To describe all of these states for all geometries, it is necessary to include orbitals beyond
the minimum set of nine, $1-5\sigma$ and $1-2\pi$, orbitals.  The anion state is the highest in
energy at infinite Rb-NH separation, but becomes the ground state at small Rb-NH separation,
and it undergoes avoided crossings with all of the other states, and so is relevant to the dynamics
on the other surfaces.  To describe the anion state, two additional orbitals are necessary:
a $\pi$ relaxation orbital that accounts for the expansion of the NH $\pi$ orbitals in the anion
state relative to the neutral states; and a correlating $\sigma^*$ orbital on the NH which increases
the Rb-NH bonding on the anion surface.  In Ref.~\cite{hutson_rbnh} a similar orbital
space was used although what we denote as an anion relaxation orbital was 
denoted by those authors as a 6$p$ orbital.

For the first step, a state-averaged
multiconfiguration SCF (MCSCF) calculation in the minimum 9-orbital space is
performed on the lowest six doublet states and the four quartets.  This 
calculation
produces a $\pi$ orbital on the NH fragment that does not account for the relaxation within
the anion state; that calculation significantly overestimates the anion state energy at small Rb-NH separations.

The next step is an all-singles-and-doubles configuration-interaction (CI) calculation on the eight doublets and four quartets in which we are
interested.  At large $R$, we must follow the anion state surface as it rises above several states that we
are not interested in.
Twelve averaged natural orbitals are adopted from this CI calculation, expanding
the orbital space to include both the $\pi$ relaxation orbitals and the $\sigma^*$ correlating
orbital.
We then perform an all-singles-and-doubles CI calculation in this space of twelve natural orbitals.   In 
both CI steps we freeze the 1-2$a'$, a.k.a. the 1-2$\sigma$, orbitals on NH, which are
basically the  N 1$s$ and 2$s$.  Among the four sets of spin and space symmetries the maximum number of configuration state functions in the CI is 1,110,312 for the doublet A$'$.

\subsection{Diabatization}

\begin{figure}[t]
\begin{center}
\begin{tabular}{cc}
\textbf{(a)} &
\resizebox{0.7\columnwidth}{!}{\includegraphics*[0.95in, 0.65in][5.6in, 3.7in]{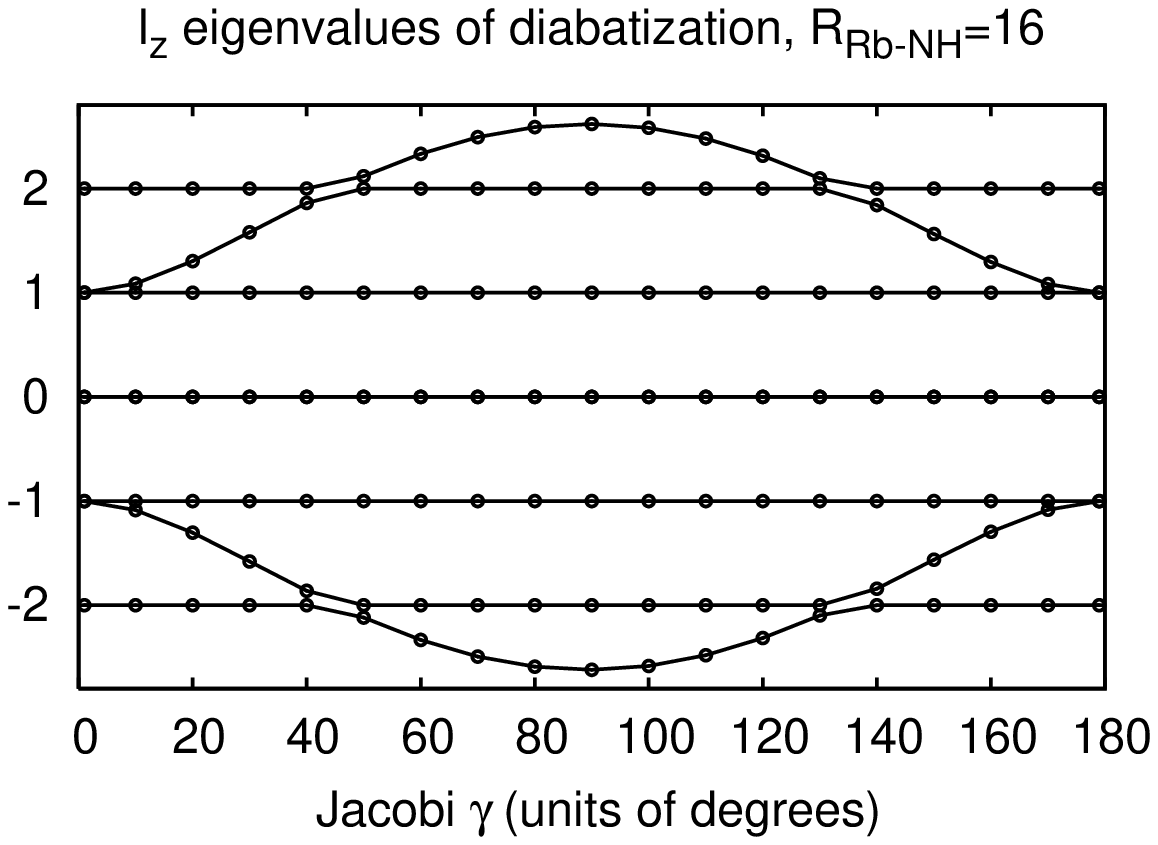}} \\
\textbf{(b)} &
\resizebox{0.7\columnwidth}{!}{\includegraphics*[0.95in, 0.65in][5.6in, 3.7in]{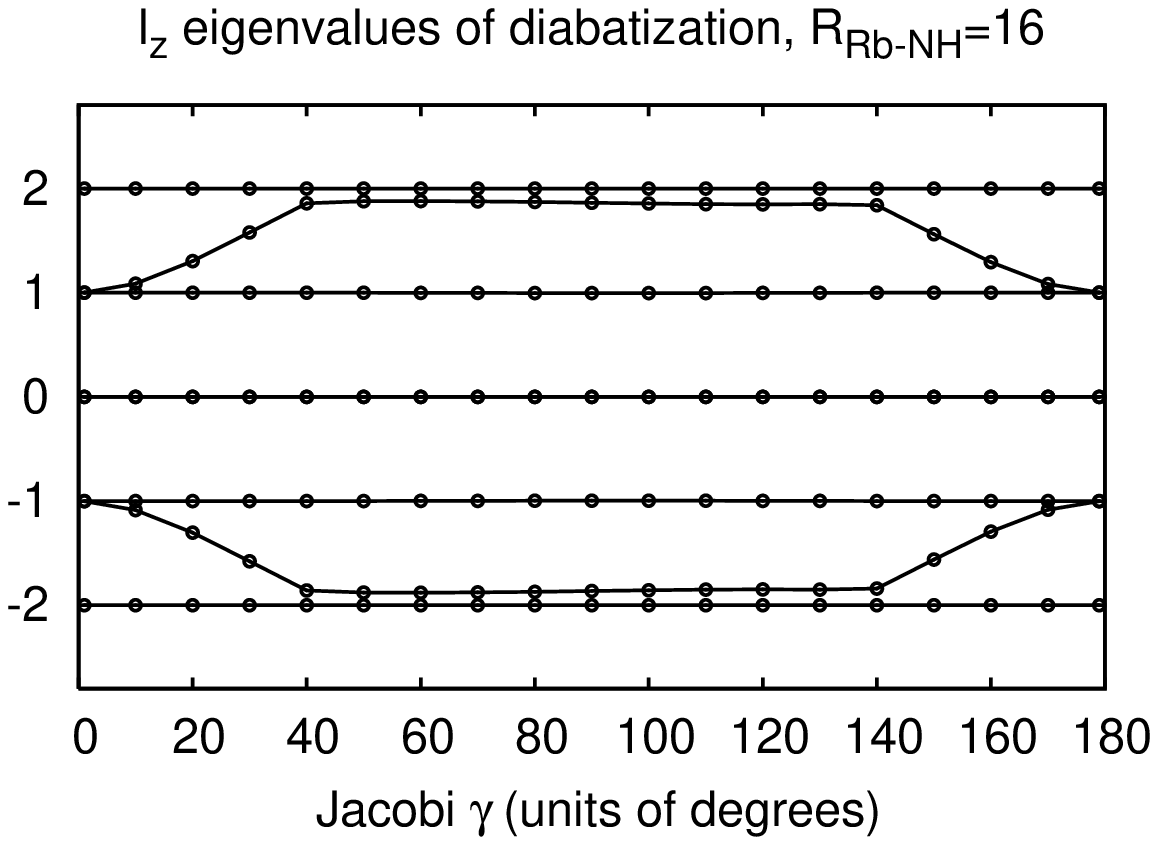}}\\
\textbf{(c)} &
\resizebox{0.7\columnwidth}{!}{\includegraphics*[0.95in, 0.65in][5.6in, 3.7in]{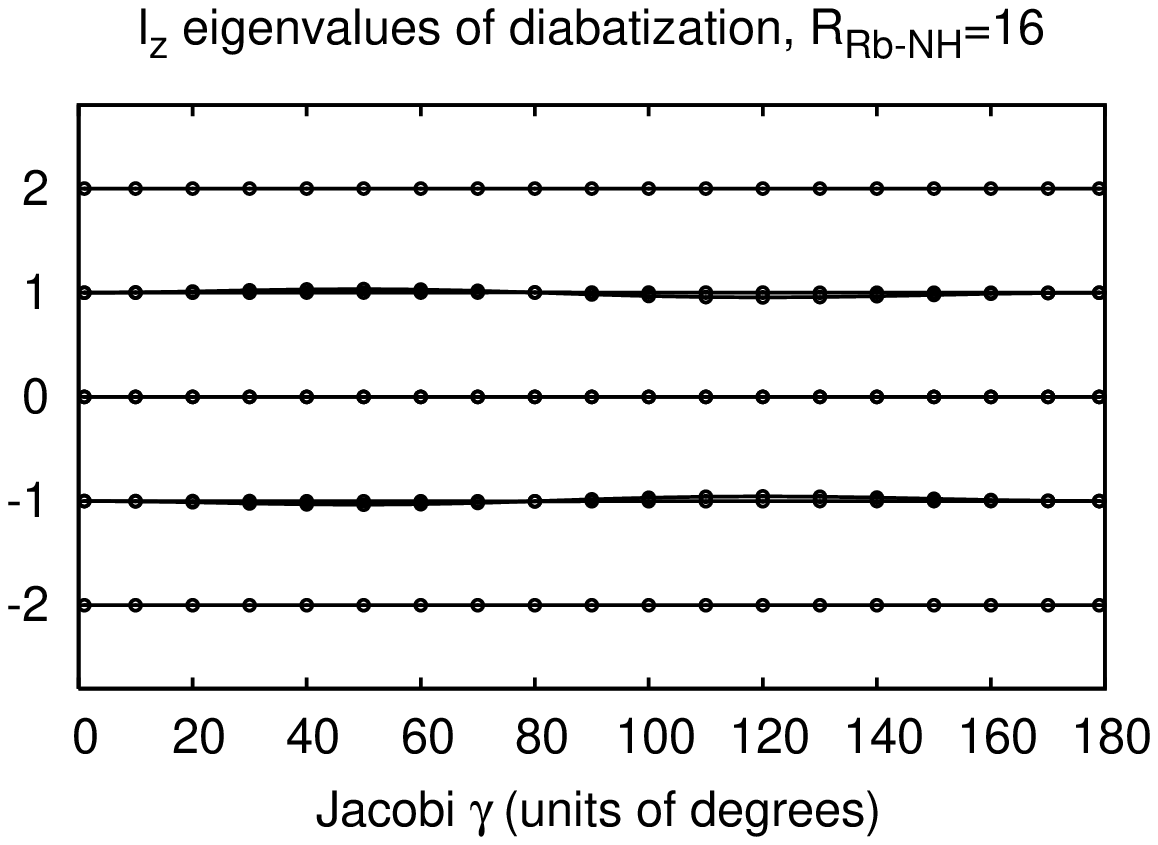}}
\end{tabular}
\end{center}
\caption{Eigenvalues and expectation values of the operator $\widehat{l_z}$ 
for diabatic states evaluated at $R_{Rb-NH}$=16.0$a_0$ as a function of the
Jacobi angle $\gamma$.  \textbf{(a)} Eigenvalues in the adiabatic basis.  \textbf{(b)} Eigenvalues in the adiabatic
basis with damping function of Eq.\ref{damping}.  \textbf{(c)}  Expectation values of 
$\widehat{l_z}$ for final diabatic states. The absolute values of these numbers provide the 
$\Lambda$ labels of the diabatic states ($^2\Sigma$, $^2\Pi$, $^2\Delta$). \label{lzfig}}
\end{figure}

At this stage of the calculation there are twelve adiabatic electronic states as functions of nuclear geometry (eight doublets and four quartets), and the spin-orbit terms have not yet been added.  A unitary transformation is now applied to these twelve states, a property-based diabatization~\cite{diareview, propertydia, dipoledia}.

The doublets are diabatized using both the electronic dipole operator and the  $z$-projection of the orbital angular momentum relative to the NH bond axis (the $\widehat{l_z}$ operator with eigenvalue $\pm\Lambda$ at linear geometry).  
The $\widehat{l_z}$ operator is first diagonalized 
in the adiabatic basis.
In order to prevent an avoided crossing at large Rb-NH
separations that would lead to a problematic mixing of our diabatic basis, 
a column and row of the $\widehat{l_z}$ matrix must be damped.  This transformation of
the $\widehat{l_z}$ operator does not affect the eigenvectors, except near the avoided
crossings.
The matrix elements of $\widehat{l_z}$ in the ground-state A$''$
row and column are thus reduced by hand beyond $R$=10.4$a_0$.  
At such large internuclear separations the lowest A$''$ state is Rb ($^2S$) + 
NH ($^3\Sigma^-$).  We reduce the matrix elements in this row and column by
a geometry-dependent factor
\begin{equation}
\forall_i \begin{split}
(\widehat{l_z})_{i,1} \rightarrow (\widehat{l_z})_{i,1}\\
(\widehat{l_z})_{1,i} \rightarrow (\widehat{l_z})_{1,i} 
\end{split}
\times
\begin{cases}
\begin{split}
& 1 \\
& \frac{10.4}{R\sin \theta}
\end{split}
\quad
\begin{split}
& R\sin \theta \le 10.4a_0  \\
\\
& R\sin \theta \ge 10.4a_0   \quad .
\end{split}
\end{cases}
\label{damping}
\end{equation}
With this damping function, the mixing between the two states NH ($^3\Sigma^-$) times Rb ($^2S$) or ($^2$P$_{\mathrm{A}''}$)
 is suppressed and the nonionic $\Pi$ state has a $\widehat{l_z}$ eigenvalue which plateaus around 1.85.

Fig.~\ref{lzfig} shows the eigenvalues of $\widehat{l_z}$ for the doublets only.  Fig~\ref{lzfig}a depicts the eigenvalues of $\widehat{l_z}$ in
the adiabatic basis at $R$=16.0$a_0$.
The $\widehat{l_z}$ eigenvalues using the damping function are plotted in
Fig~\ref{lzfig}b.  The final expectation values of $\widehat{l_z}$ in our diabatic basis
are shown in Fig.~\ref{lzfig}c; these values are nearly integers, which permits the labeling of these states by the $\Lambda$ quantum number labels.  We have a total of two $^2\Sigma$, two $^2\Pi$, and one $^2\Delta$, and two $^4\Sigma$ and one $^4\Pi$ diabatic state.

\begin{figure}
\begin{center}
\begin{tabular}{cc}
a) &
\resizebox{0.8\columnwidth}{!}{\includegraphics*[0.80in, 0.6in][5.63in, 3.7in]{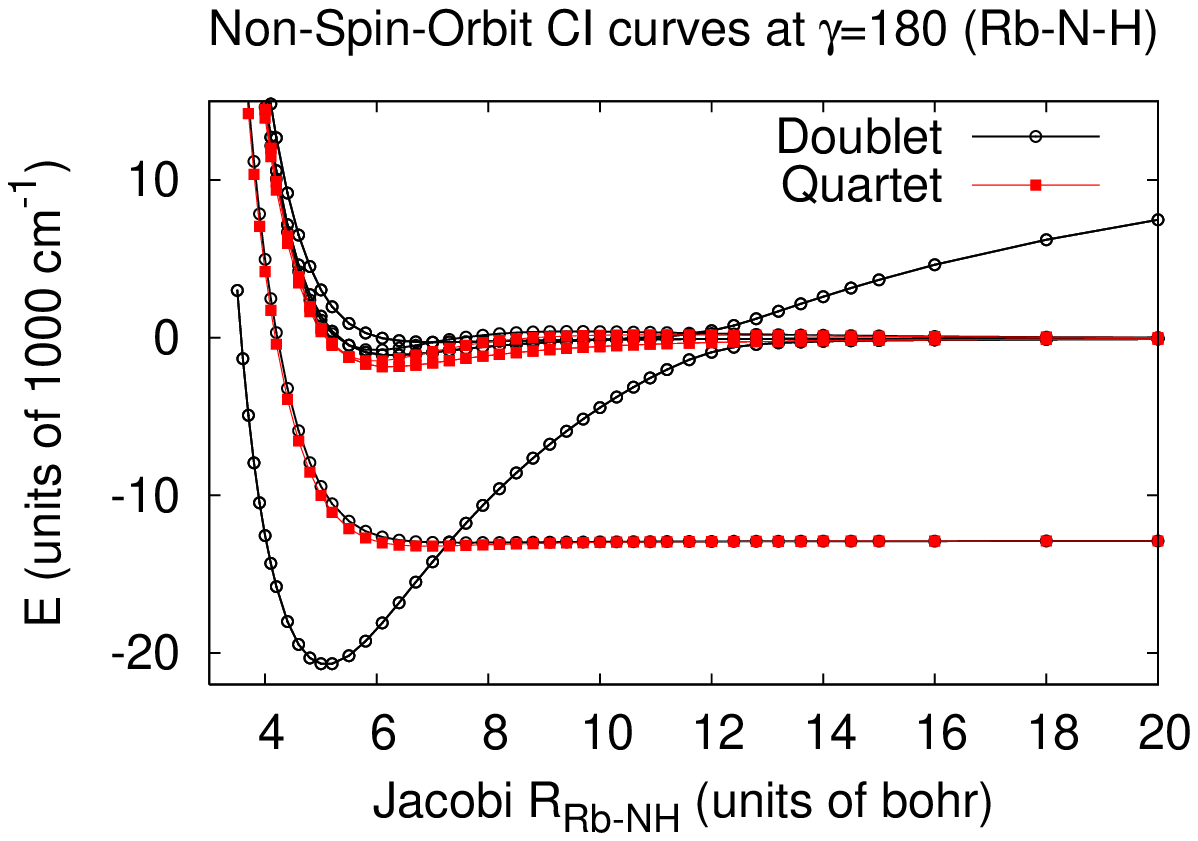}} \\
b) &
\resizebox{0.75\columnwidth}{!}{\includegraphics*[0.78in, 0.6in][5.52in, 4.1in]{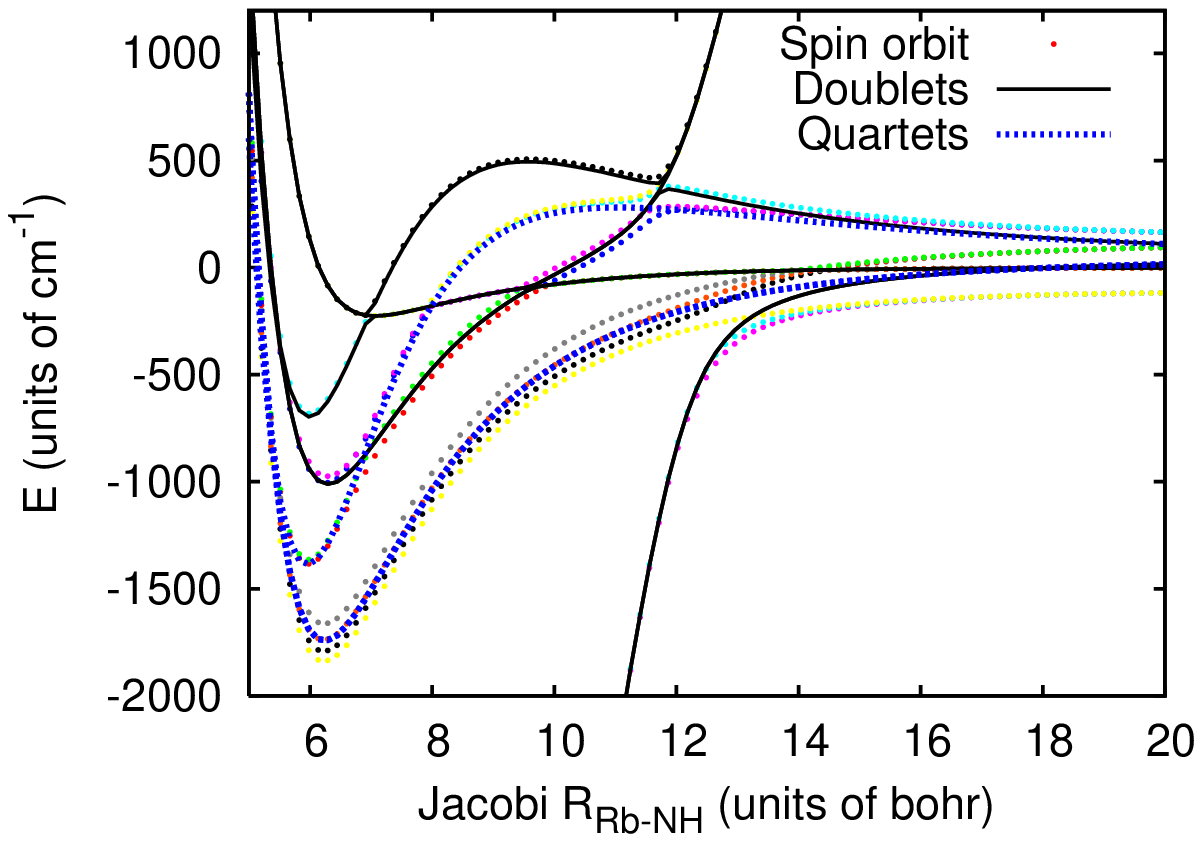}} \\
\end{tabular}
\end{center}

\caption{\textbf{a)} Non-spin-orbit CI results at $\gamma=180^\circ$.  
\textbf{b)}
Final spin-orbit surfaces used in study, obtained by adding the
spin-orbit term by hand to the diabatic basis, $\gamma$=180$^\circ$. 
\label{finalfig}}
\end{figure}

The electronic dipole operator is used to separate the two $^2\Pi$ states.  These correlate with the anion NH$^-$ ($^2\Pi$) + Rb$^+$ state and with NH
($X ^3\Sigma^-$) + Rb ($^2P$) states at infinity; at small Rb-NH separations the former
is the lower electronic state and at large separations the latter is.
The interaction between these states is such
that they form an avoided crossing of width approximately 1000 cm$^{-1}$ around $R_{Rb-NH}$ = 11$a_0$.
The situation is similar to that of RbOH~\cite{hutson_rboh_prospects, hutson_rboh}.
 Diagonalization of the electronic dipole operator in the direction of the Jacobi 
vector $\vec{R}$ is carried out in the $\Pi^+$ and $\Pi^-$ spaces separately, and the electronic Hamiltonian is diagonalized in the $^2\Sigma$ space.
The quartets are diabatized by diagonalization of $\widehat{l_z}$, followed by diagonalization of the electronic Hamiltonian in
the $^4\Sigma$ space.  The expectation values of $\widehat{l_z}$ in our final diabatic basis are all near integers.

\subsection{Adding spin-orbit terms}

At this point, then, we have a diabatic basis in which each member is labeled by a given projection of electronic angular momentum about the NH bond axis, and in which the spin-orbit part of the electronic Hamiltonian has not been included.  Multiplying each of these electronic basis functions by all possible spin wavefunctions, we arrive at the full 32-microstate basis.  Assuming that our diabatic basis function labels are indeed good quantum numbers, the addition of the spin-orbit terms is straightforward, requiring only the enforcement of the Condon-Shortley phase convention among the members of the diabatic basis, 
and the spin-orbit terms have no dependence upon the nuclear geometry.  The discussion below verifies that this approximation faithfully reproduces the true eigenvalues.

After adding these spin-orbit terms we shift the delta diabatic potential 
energy surfaces slightly (by tens of wavenumbers) such that the asymptotes 
coincide with their physical values.  A unitary 
transformation now yields the two uncoupled 16$\times$16 blocks each 
containing one member of every Kramers doublet.  These blocks are complex conjugates of one another; it is only necessary to choose one of these to use as the Born-Oppenheimer electronic Hamiltonian for total angular momentum $J=0$.  Extensions to include the electronic angular momentum (the Renner-Teller and spin Renner-Teller effect) and nonzero $J$ will instead use the real-valued 32$\times$32 representation.

\subsection{Description of the surfaces}

Results of the non-spin-orbit CI for linear Rb-N-H geometry are shown in Fig.~\ref{finalfig}a.  The ground state in the asymptotic region is Rb ($^2S$) + NH ($X ^3\Sigma^-$) with an asymptote of -55.1993162 hartree.

The $^2\Pi$ anion surface comes sweeping down, with a minimum at 5.1192$a_0$ of -55.2346523 hartree, giving a depth of 0.949eV.  In comparison, Ref.~\cite{hutson_rbnh} found a well depth of 1.372eV at $R$=4.911$a_0$, for a NH bond length of 1.948$a_0$.  It crosses the collection of nearly degenerate channels at approximately 12.0$a_0$, and crosses the surface that is asymptotically the ground state at 7.3029$a_0$ and at an energy of -55.1997712 hartree.  In comparison, Ref.~\cite{hutson_rbnh} found the crossing at 7.163$a_0$, also for a NH bond length of 1.948$a_0$.

The final results including the spin-orbit interaction, 
obtained by adding the spin-orbit terms by hand to the
diabatic Hamiltonian, are plotted in Fig.~\ref{finalfig}b, in a closer view.  The $^2\Delta$ state is visible as the flattest line, staying near zero wavenumbers until around 10$a_0$ on the scale of the figure.  At the far edge of the figure  the surfaces have separated into the Rb ($^2$S) $\times$ NH ($^1\Delta$) asymptote, in the middle, with Rb ($^2$P$_{3/2}$) and  Rb ($^2$P$_{1/2}$) times NH ($X ^3\Sigma^-$) above and below, respectively.

\begin{figure}
\begin{center}
\resizebox{0.9\columnwidth}{!}{\includegraphics*[0.7in, 0.63in][5.7in, 4.1in]{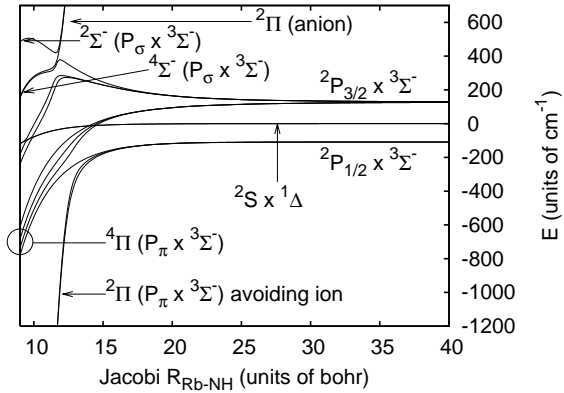}} 
\end{center}
\caption{Long-range behavior of surfaces, $\gamma=180^\circ$. \label{finalfig2}}
\end{figure}

Components of the upper $^2$P$_{3/2}$ state split into sets following either the $^{2,4}\Sigma$ surface, which rises from its asymptote inward, or the $^{2,4}\Pi$ surfaces.  These latter surfaces drop in energy going inward from their asymptote -- the quartet less so, whereas the $^2\Pi$ surface undergoes an avoided crossing with the anion $^2\Pi$ surface, dropping steeply downward from the top of the figure.  Inward of this avoided crossing around 12.0$a_0$, the $^2\Pi$ (Rb $^2$P $\times$ NH $X ^3\Sigma^-$) state is found dropping downward with decreasing $R$ from the avoided crossing and crosses the $^2\Delta$ surface around 9.0$a_0$.  There are other crossings farther in, too numerous to mention.

\begin{figure}
\begin{center}
\resizebox{0.9\columnwidth}{!}{\includegraphics*[0.8in,0.6in][5.4in,4.1in]{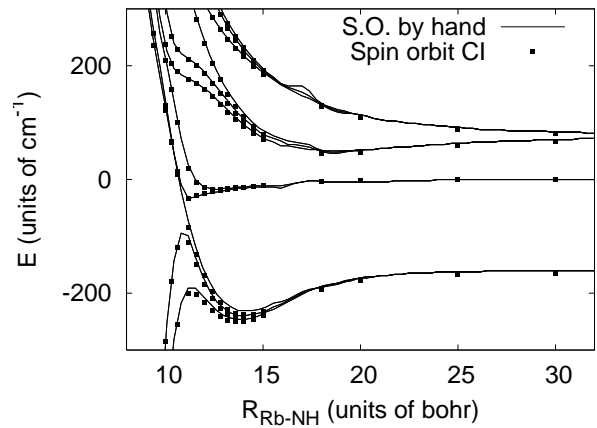}}
\end{center}
\caption{Spin-orbit results using a smaller, 9 orbital valence space CI, $\gamma$=30$^\circ$.  Spline-interpolated surfaces calculated from diabatic states adding the
spin-orbit
interaction by hand (lines) compared with
spin-orbit CI results (points). The asymptotes are not correct for this test calculation.  \label{zoom30}}
\end{figure}

The long range behavior of the
surfaces is shown in Fig.~\ref{finalfig2}.  This behavior is governed by the long-range interactions: as the NH has a dipole moment, these are dipole-quadrupole with $\frac{1}{r^4}$ power law for the Rb ($^2$P) fragment and dipole-induced dipole with $\frac{1}{r^6}$ power law for the $^2$S fragment.  The dipole-quadrupole interaction, acting on the Rb $^2$P $\times$ NH $^3\Sigma^-$ channels, is stronger than the dipole-induced dipole interaction for the $^2\Delta$ channel and therefore the crossings are determined by the former.  At linear Rb-N-H geometry, this interaction splits the $^2$P state into the sigma component at higher energy and the pi component  at lower energy, which behavior we described above for the results in Fig.~\ref{finalfig}b.  It is responsible for the crossings around 14.7$a_0$.  We examined these crossings for various Jacobi angles $\gamma$; for all angles the crossing is apparently too sharp to provide a mechanism for coupling between the $^2\Delta$ and the components of the Rb ($^2P_{3/2}$) state that correlate with $\Pi$ symmetry farther in.  Results below seem to indicate that this is indeed the case.  The relevant crossings for collisions at low energy are those visible in Fig.~\ref{finalfig}b.

\subsection{Verification of treatment of spin-orbit effect}

The spin-orbit calculation using the COLUMBUS program
was prohibitively large to perform in the full orbital
space.  We thus perform the
spin-orbit calculation only at the first, nine-orbital CI step, as described 
above.  Results of this calculation are presented in Fig.~\ref{zoom30}.
These results verify that addition of the spin-orbit terms by hand to the diabatized Hamiltonian from the
non-spin-orbit CI calculation reproduces the results of the full spin-orbit CI.  (The asymptotes are not correct for the test calculation shown in this figure.)

\section{Scattering calculation}

We calculate the quantum nuclear dynamics on the coupled set of diabatic
potential energy surfaces for total angular momentum $J=0$.
The standard~\cite{petrongolo, hd2}
BF Hamiltonian for $R$ times the wavefunction,
keeping $r$ fixed, in Jacobi coordinates is

\begin{equation}
\begin{split}
H& =-\frac{1}{2\mu_R} \frac{\partial^2}{\partial R^2}
+ B_r \hat{j}^2 + \frac{1}{2\mu_R R^2}\hat{j}^2  + V(R,r,\gamma)  \\
\hat{j}^2 & =  -\left(\frac{1}{{\mathrm{sin}}
 (\gamma)}\frac{\partial}{\partial \gamma} {\mathrm{sin}}
 (\gamma) \frac{\partial}{\partial \gamma} \right) \qquad ,\\
\label{hamiltonian}
\end{split} 
\end{equation}
where $\mu_R$ is the reduced mass in that degree of freedom;
$B_r$ is the rotational constant of NH, taken to be 16.699cm$^{-1}$~\cite{herz}, the
value for the ground electronic state;
and $V$ is the matrix representation of the electronic Hamiltonian in the diabatic basis.

The scattering calculations employ the R-matrix propagator technique of Baluja, Burke and Morgan~\cite{propagator}.
Our implementation adopts the discrete variable representation (DVR)~\cite{dickcert,lhl}, with
the Legendre DVR~\cite{cor92:4115} in $\gamma$ -- with 80 points -- and the Gauss-Lobatto DVR~\cite{femdvr}
in $R$, with 6 points per element, 960 elements, from 3.25$a_0$ to 43.25$a_0$.
The propagation covers one element at a time.
For each diabatic electronic state, the basis in $\gamma$ is contracted by calculation of an adiabatic basis in $\gamma$ as a function of the scattering coordinate $R$ and include 42 adiabatic states in $\gamma$ per diabatic electronic state.  Slow variable discretization (SVD)~\cite{svd} efficiently accounts for the coupling between these surfaces nonadiabatic in $R$.

\begin{figure}[t]
\begin{center}
\begin{tabular}{c}
\resizebox{0.75\columnwidth}{!}{\includegraphics*[0.8in, 0.65in][5.6in, 11.5in]{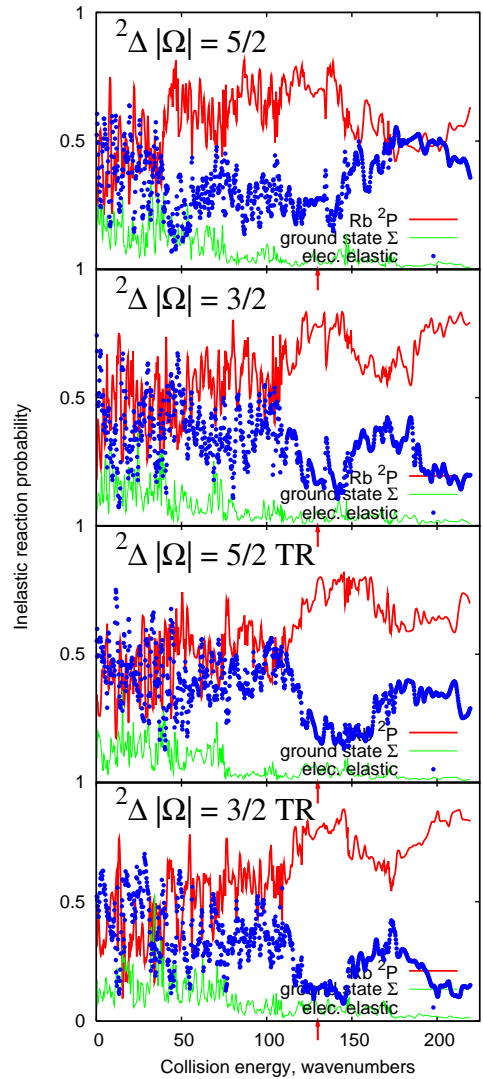}} \\
\end{tabular}
\end{center}
\caption{Results of scattering calculations: the probability per collision for a transition from the lowest rovibrational state of $^2\Delta$ (NH $^1\Delta$
$\times$ Rb $^2S$) to other electronic states as a function of collision energy, for $J=0$.  The four unique initial states are described in the text.  The thick red (dark) line is the probability for transition to any of the Rb ($^2$P) channels; the thin green (light) line is that for a transition to any of the ground $^{2,4}\Sigma$ channels; the dots mark the probability of staying in the $^2\Delta$ channel space.  The arrow marks the energy at which the $^2$P$_{3/2}$ channels bcome open. \label{scatfig}}
\end{figure}

Results of the scattering calculation are shown in Fig.~\ref{scatfig}.  The total probabilities for transitions from the ground rovibrational state of NH
($^1\Delta$) plus Rb ($^2S$) to the other electronic states are plotted as functions
of collision energy in the incident channel.  Two degenerate $^2\Delta$ channels in the calculation 
have $\vert\Omega\vert$ = $\frac{3}{2}$ or $\frac{5}{2}$,
 but the method permits the evaluation of four independent cross sections, for both the calculated 
process and the time-reversed process corresponding to the the other 16$\times$16
Kramers doublet block (i.e., complex conjugating the electronic hamiltonian in the diabatic basis).  
The four corresponding sets of scattering amplitudes correspond to  two rows and two columns of the calculated S-matrix.  The S-matrix cannot be chosen
to be symmetric because we have excluded the time-reversed orthogonal complement of our sixteen states, the sixteen 
states of the other Kramers doublet block.   The label ``TR'' in Fig.~\ref{scatfig}
denotes the time-reversed partner.

These results indicate that the cross section for the quenching reaction is indeed large in the 
present treatment, and it is in fact comparable to the unitarity limit.  In contrast, the 
cross section to the ground $^{2,4}\Sigma$ channels is significant for low collision energy but is 
at least an order of magnitude below the quenching reaction cross section at energies above the Rb $^2$P$_{3/2}$ 
threshold, which is marked with an arrow on the abscissa of Fig. \ref{scatfig}.  A 
qualitative change in the branching ratios becomes evident near this energy, 128 wavenumbers; the lower three 
panels of Fig.~\ref{scatfig} indicate that for collisions in those incident channels, the proportion of
Rb ($^2P$) + NH ($^3\Sigma^-$) to Rb ($^2S$) + NH ($^1\Delta$) produced increases at higher energy.
However, the correlation of features such as the change in this proportion with the opening of the
$^2P_{3/2}$ channels is not perfect, and there is the possibility that it derives from the opening of different channels
(for instance, the Rb ($^2S$) + NH ($^1\Delta \ j=2$) channel at 100 wavenumbers, or  Rb ($^2P_{1/2}$) + NH ($^3\Sigma^- \ j=3$) at 91 wavenumbers), or another mechanism altogether.  
A fuller characterization of these results will be presented in a forthcoming publication.

\begin{figure*}
\begin{center}
\begin{tabular}{ccc}
\resizebox{0.58\columnwidth}{!}{\includegraphics*[1.85in, 1.6in][4.6in,3.6in]{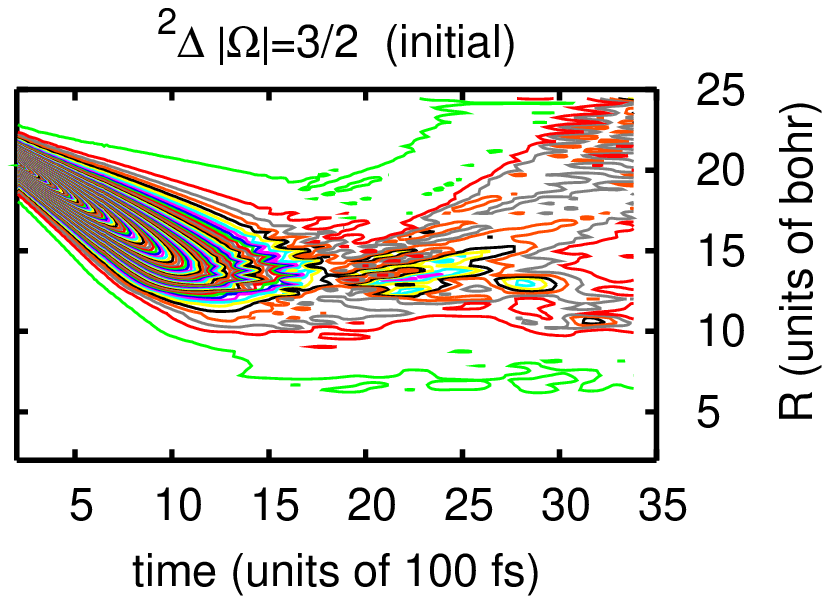}} &
\resizebox{0.58\columnwidth}{!}{\includegraphics*[1.85in, 1.6in][4.6in,3.6in]{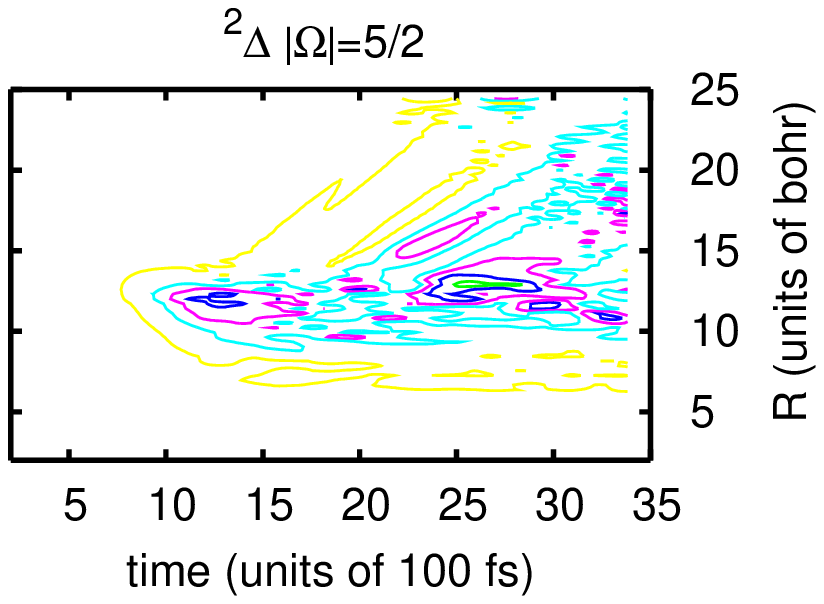}} &
\resizebox{0.70\columnwidth}{!}{\includegraphics*[1.85in, 1.6in][5.2in,3.6in]{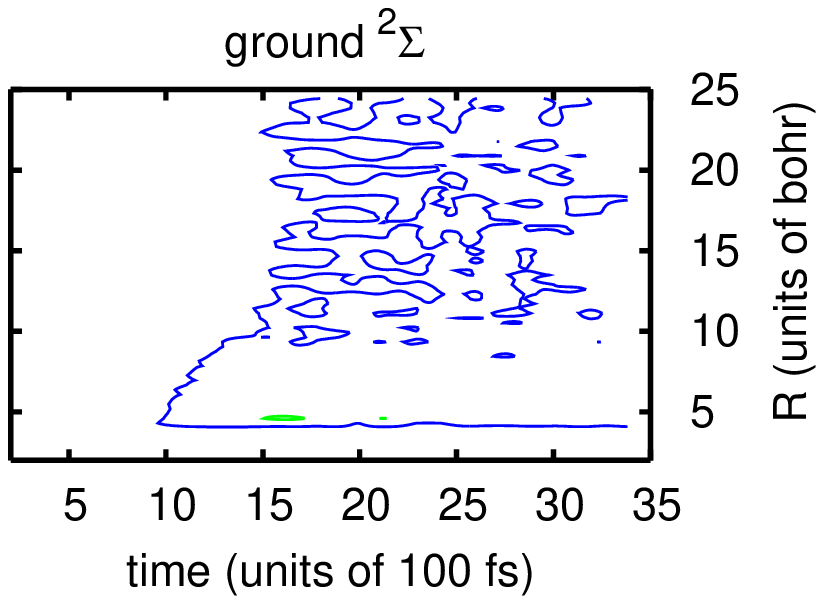}} \\
\resizebox{0.58\columnwidth}{!}{\includegraphics*[1.85in, 1.6in][4.6in,3.6in]{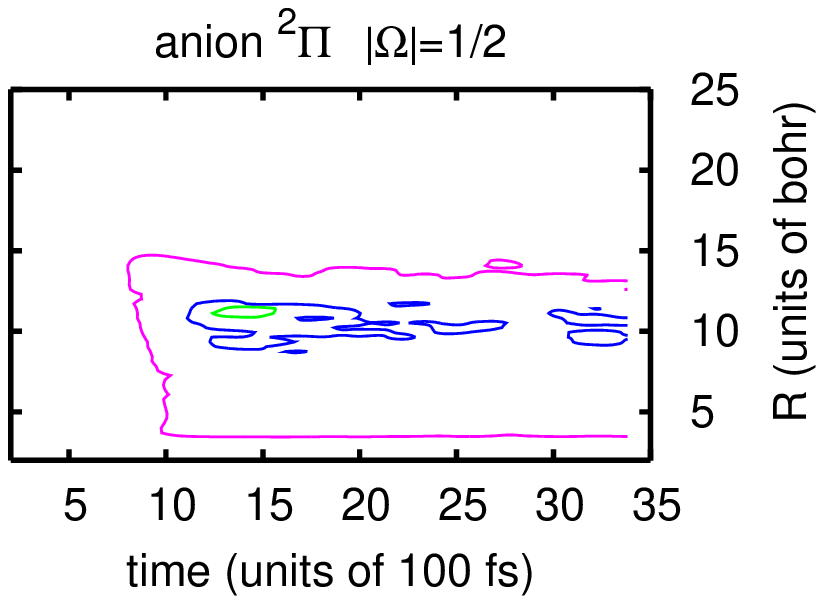}} &
\resizebox{0.58\columnwidth}{!}{\includegraphics*[1.85in, 1.6in][4.6in,3.6in]{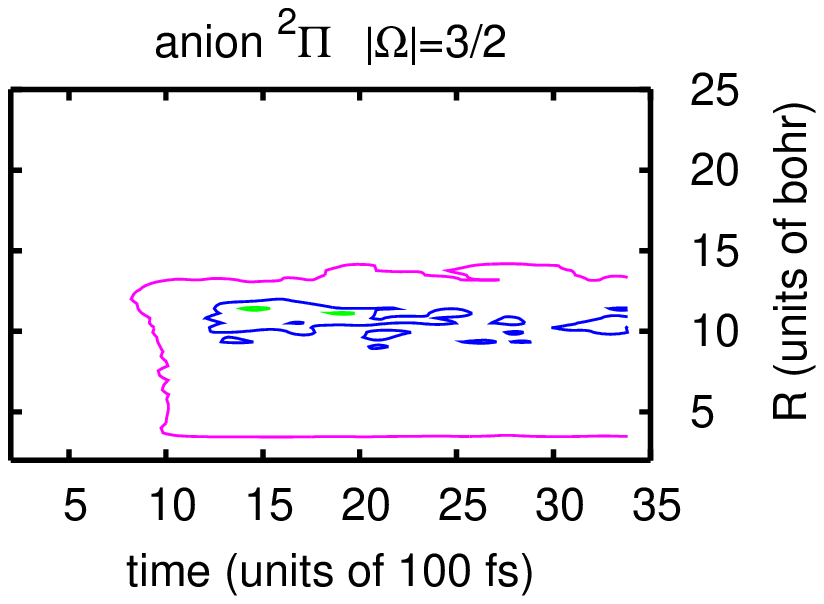}} &
\resizebox{0.70\columnwidth}{!}{\includegraphics*[1.85in, 1.6in][5.2in,3.6in]{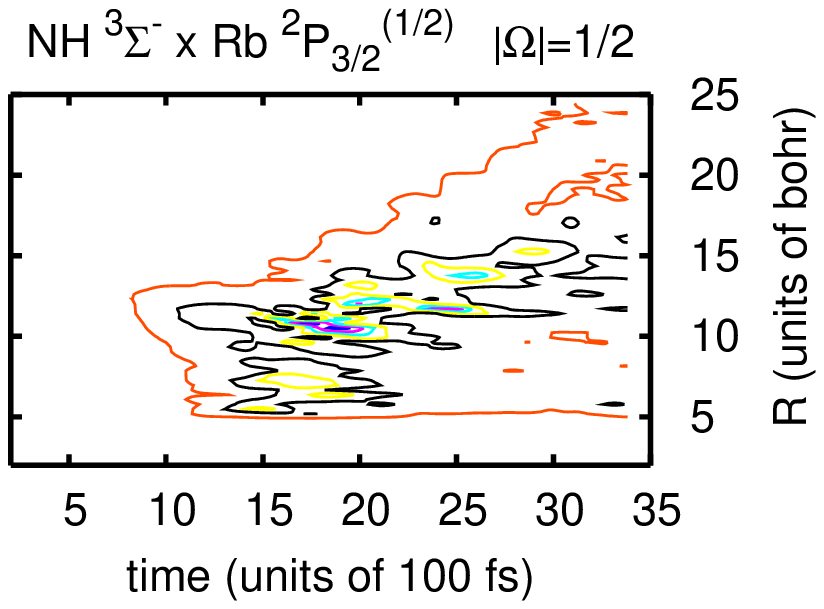}} \\
\resizebox{0.58\columnwidth}{!}{\includegraphics*[1.85in, 1.6in][4.6in,3.6in]{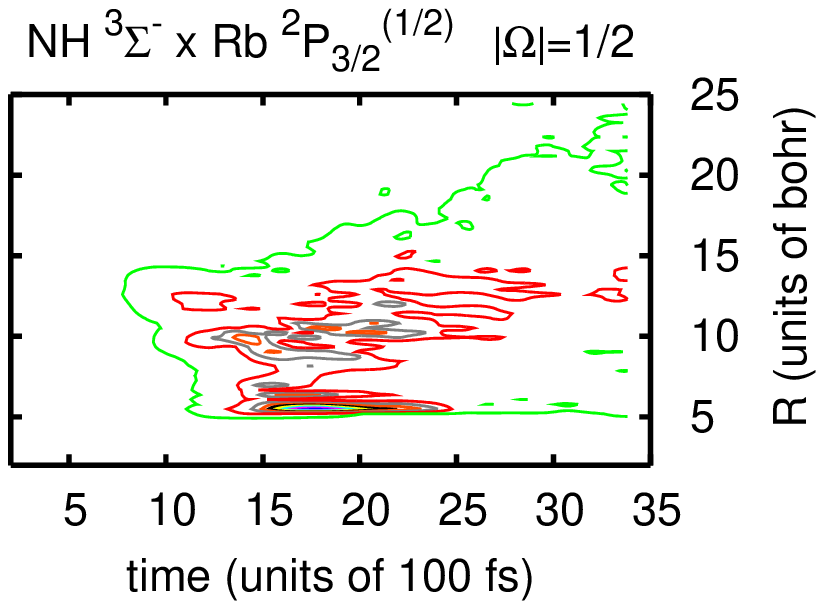}} &
\resizebox{0.58\columnwidth}{!}{\includegraphics*[1.85in, 1.6in][4.6in,3.6in]{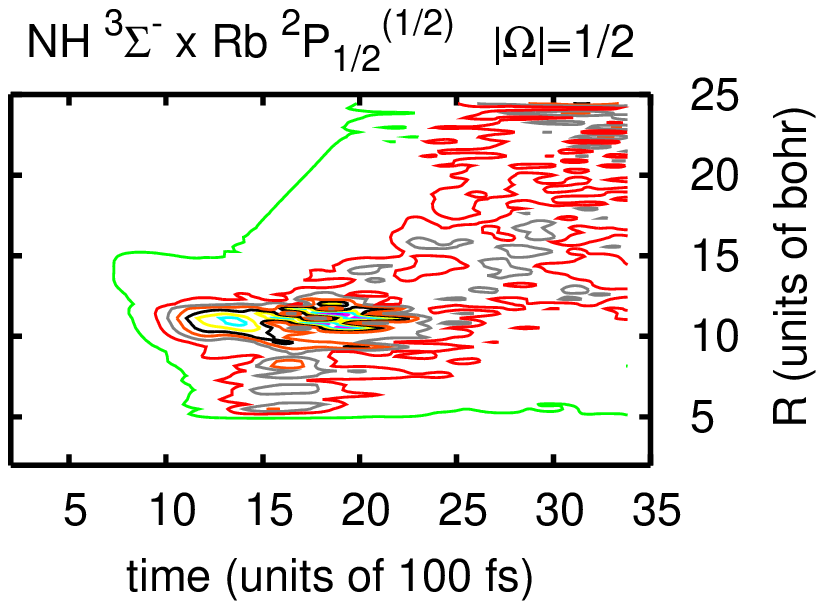}} &
\resizebox{0.70\columnwidth}{!}{\includegraphics*[1.85in, 1.6in][5.2in,3.6in]{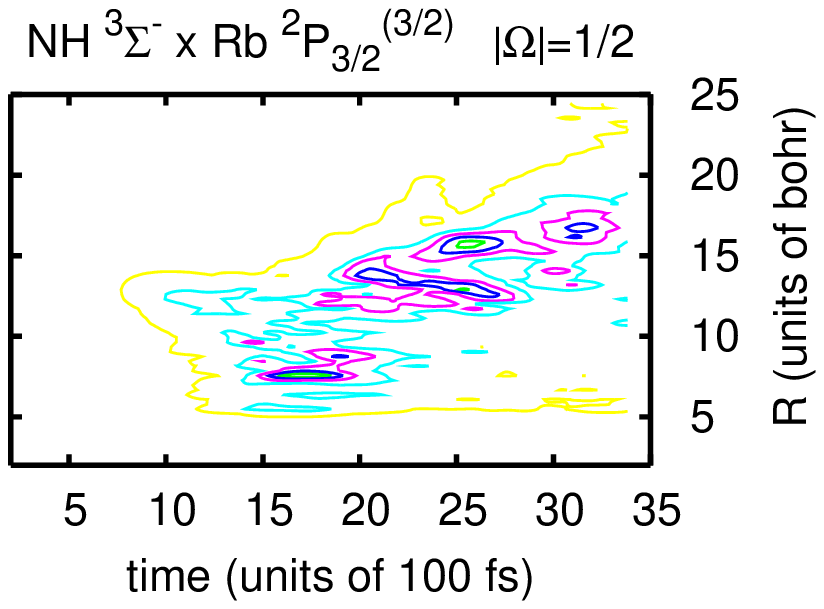}} \\
\resizebox{0.58\columnwidth}{!}{\includegraphics*[1.85in, 1.6in][4.6in,3.6in]{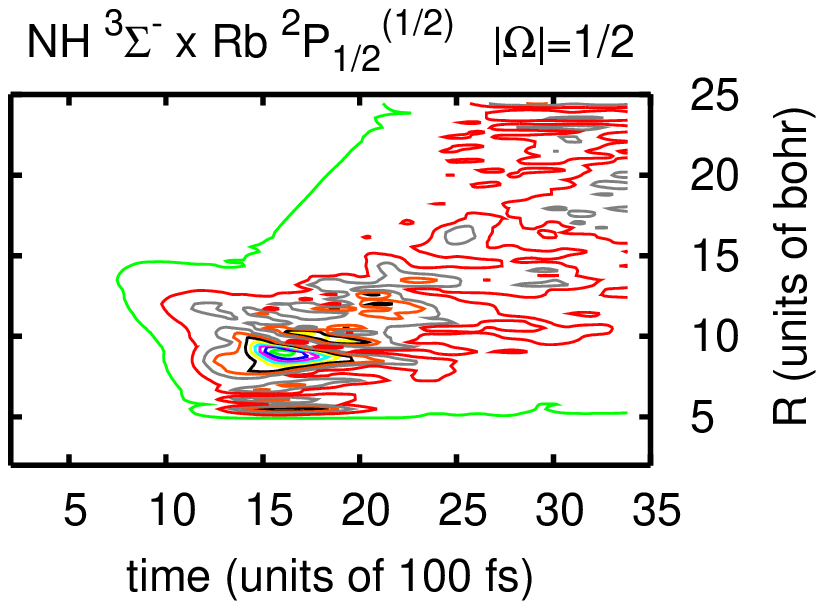}} &
\resizebox{0.58\columnwidth}{!}{\includegraphics*[1.85in, 1.6in][4.6in,3.6in]{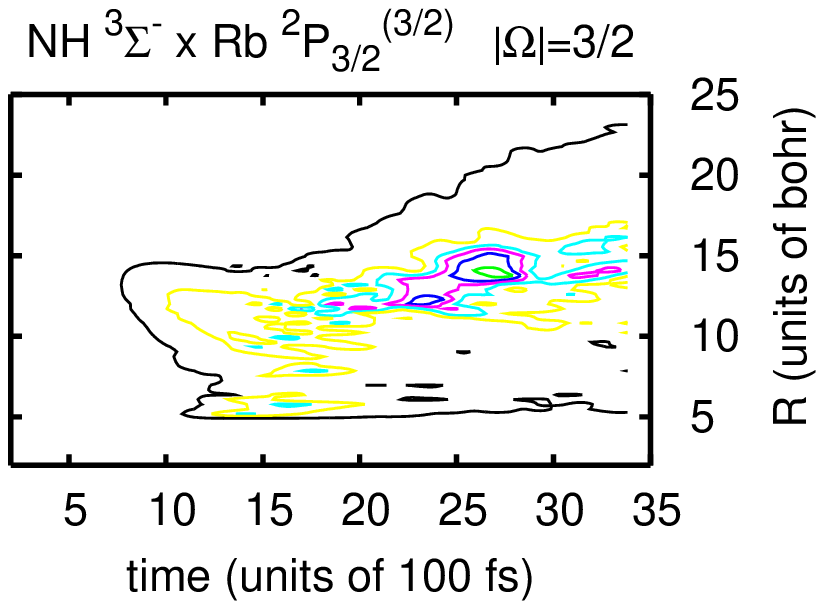}} &
\resizebox{0.70\columnwidth}{!}{\includegraphics*[1.85in, 1.6in][5.2in,3.6in]{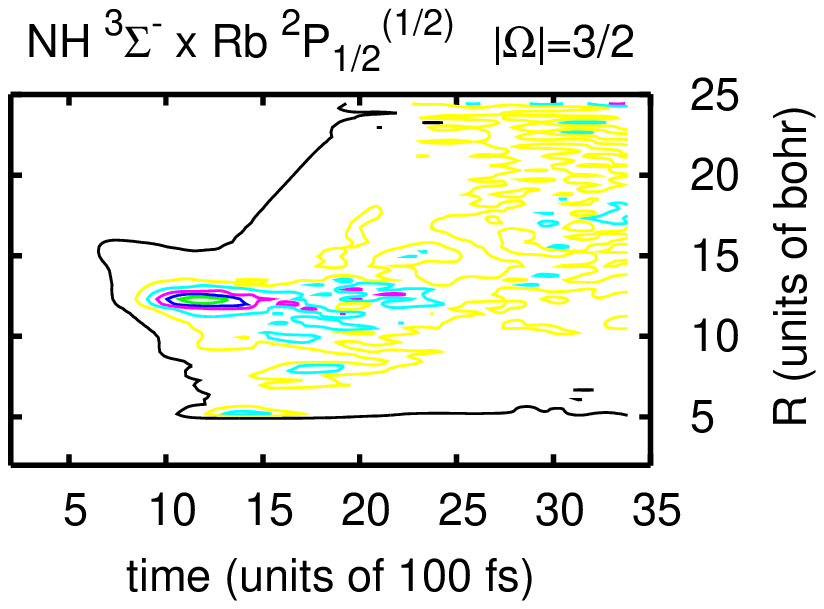}} \\
\resizebox{0.58\columnwidth}{!}{\includegraphics*[1.85in, 1.0in][4.6in,3.6in]{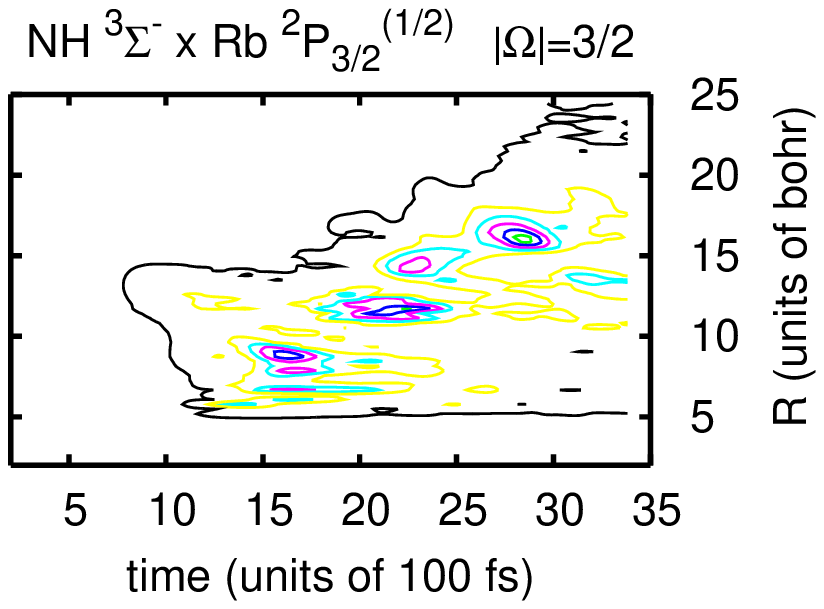}} &
\resizebox{0.58\columnwidth}{!}{\includegraphics*[1.85in, 1.0in][4.6in,3.6in]{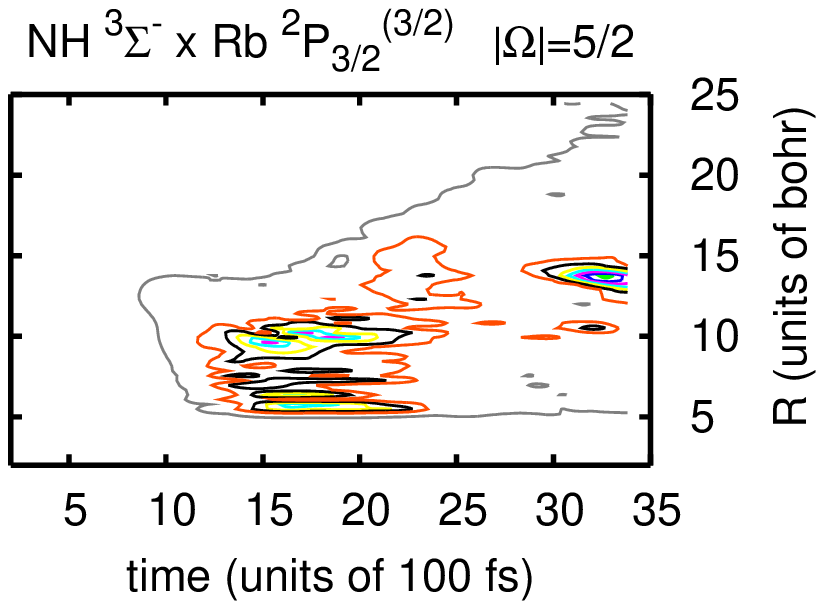}} &
\resizebox{0.70\columnwidth}{!}{\includegraphics*[1.85in, 1.0in][5.2in,3.6in]{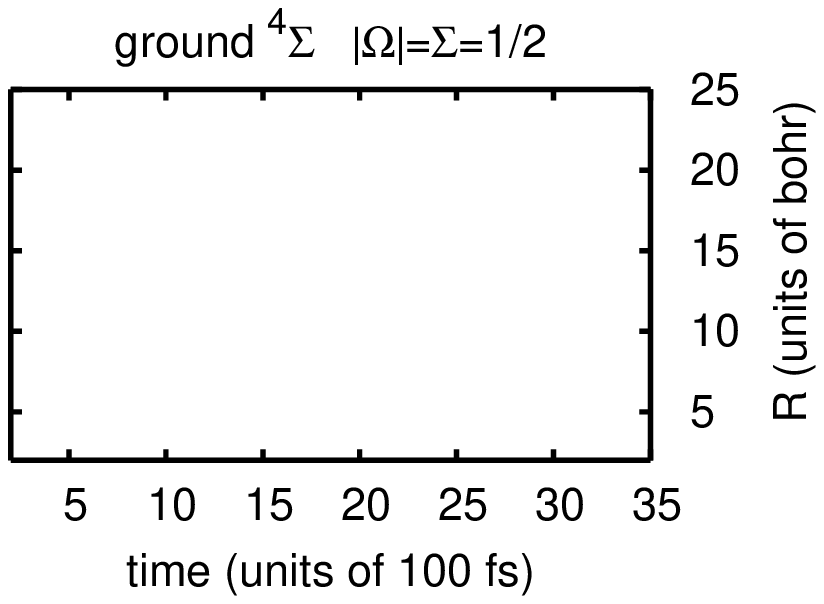}} \\
\end{tabular}
\end{center}
\caption{Results of time-dependent wavepacket propagation described in the text.  The density, in arbitrary units,
 integrated over $\gamma$, on
each electronic channel is shown as a function of the time of propagation and the value of $R$.  The results for
the last $^4\Sigma$ state are the same as the one shown, i.e., insufficient density to be visible on the plot.
The first contour is at $\frac{1}{10}$th the value of the second one, and the spacing is linear.  The states are in the same order (left to
right, top to bottom) as in Table~\ref{so_diatable}.\label{propfig}}
\end{figure*}

\section{Time-dependent calculations}

\begin{figure*}
\begin{center}
\begin{tabular}{cc}
\resizebox{1.0\columnwidth}{!}{\includegraphics{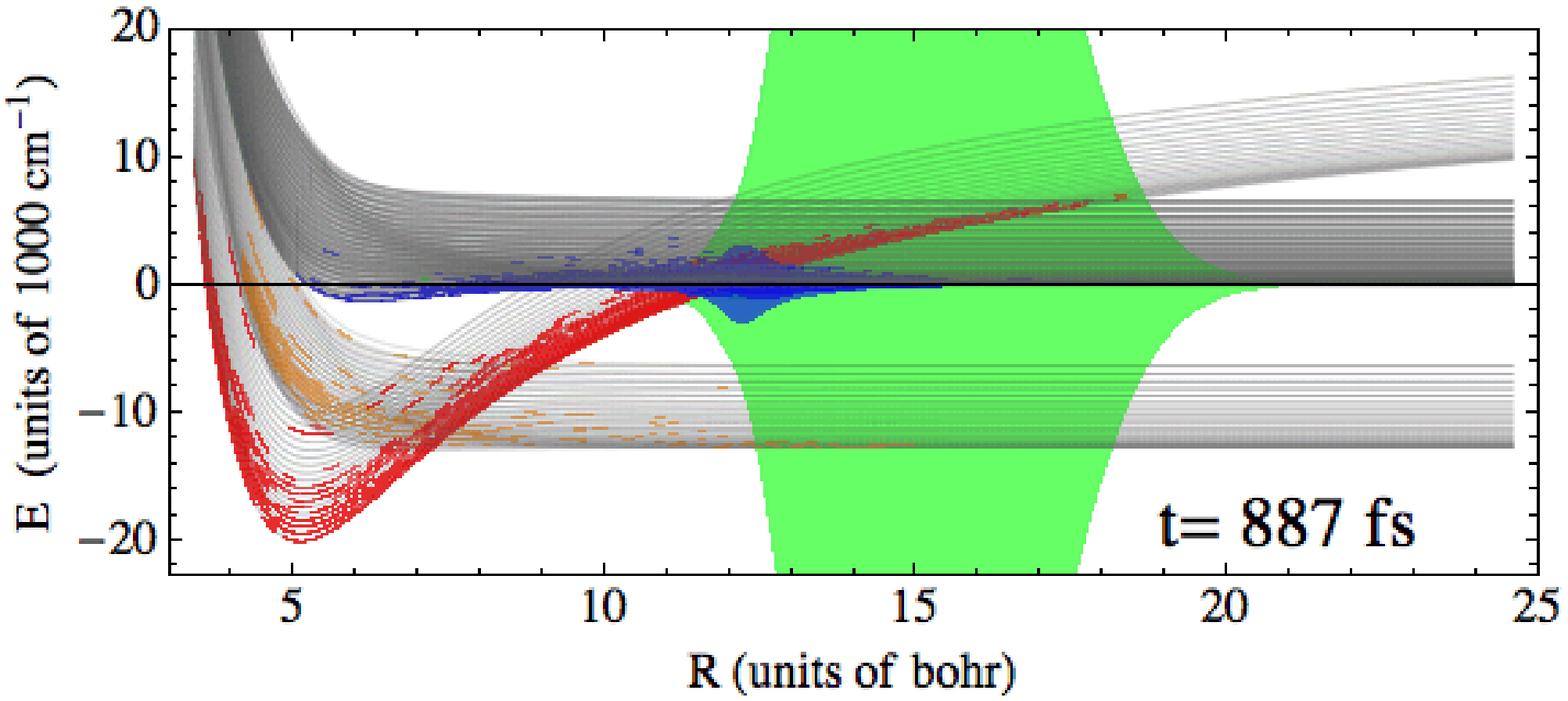}} &
\resizebox{1.0\columnwidth}{!}{\includegraphics{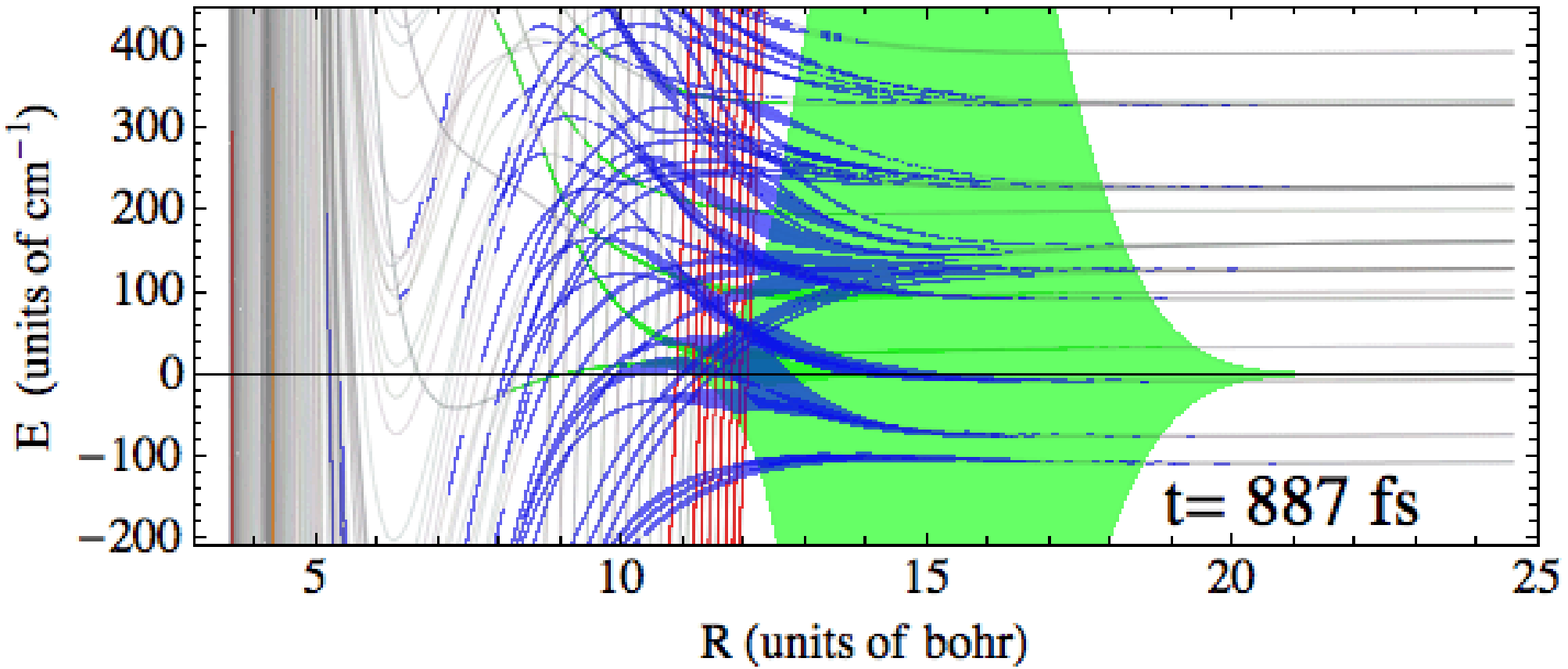}} \\
\resizebox{1.0\columnwidth}{!}{\includegraphics{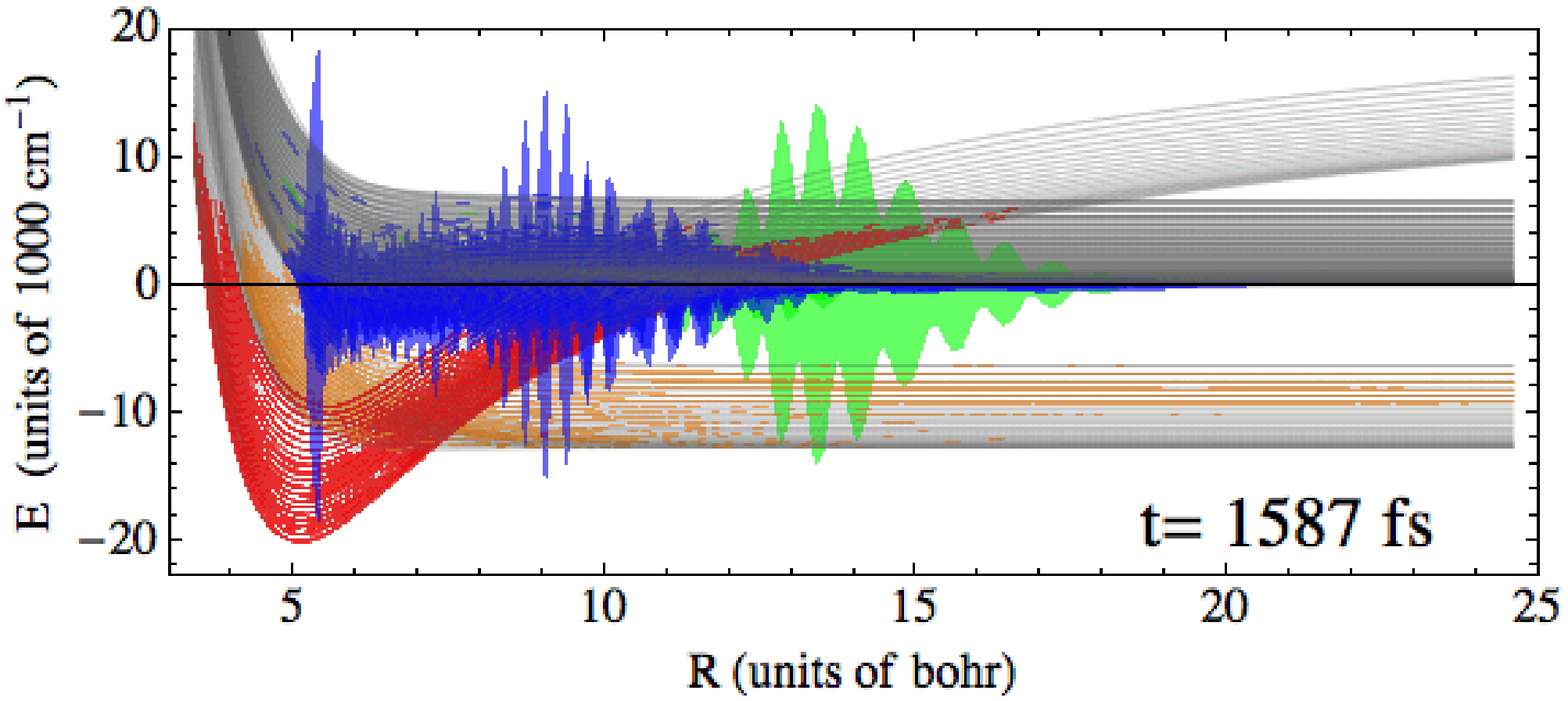}} &
\resizebox{1.0\columnwidth}{!}{\includegraphics{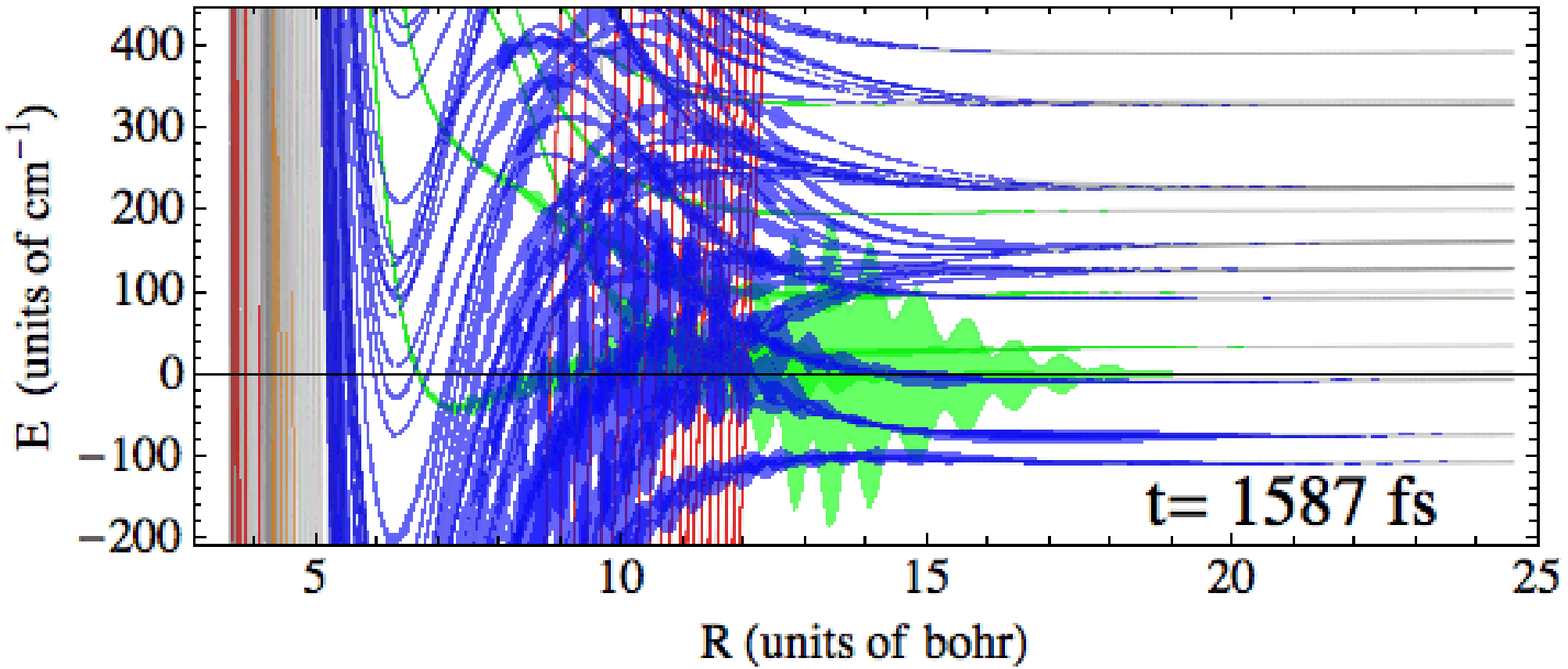}} \\
\resizebox{1.0\columnwidth}{!}{\includegraphics{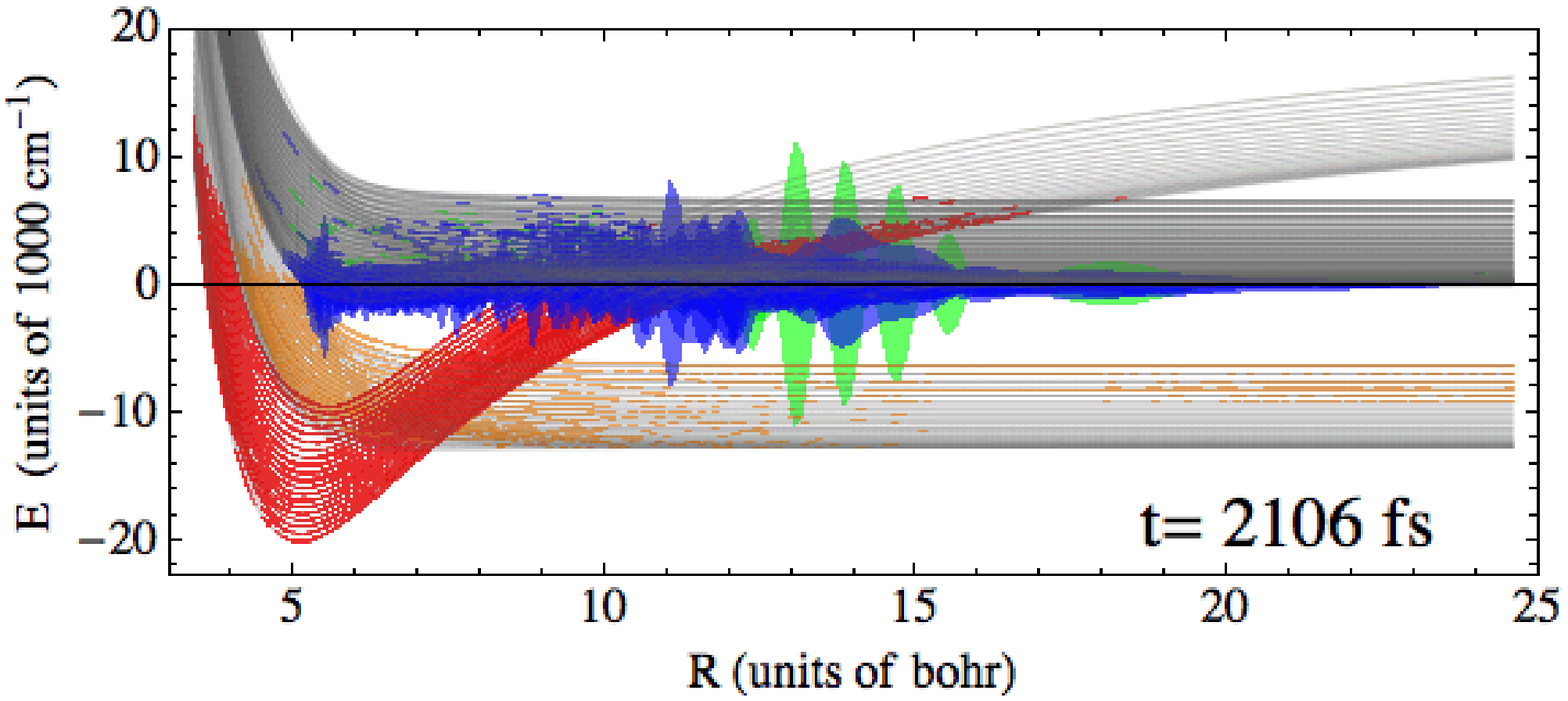}} &
\resizebox{1.0\columnwidth}{!}{\includegraphics{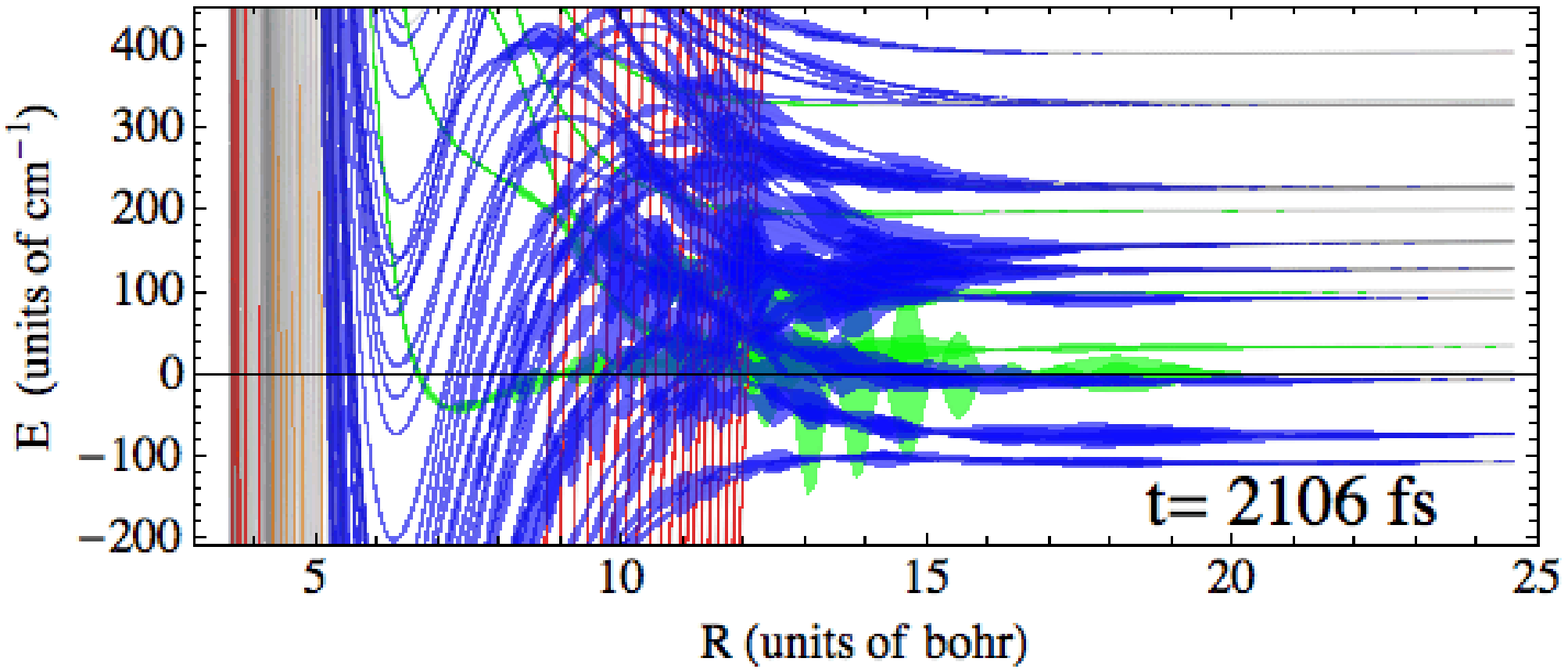}} \\
\resizebox{1.0\columnwidth}{!}{\includegraphics{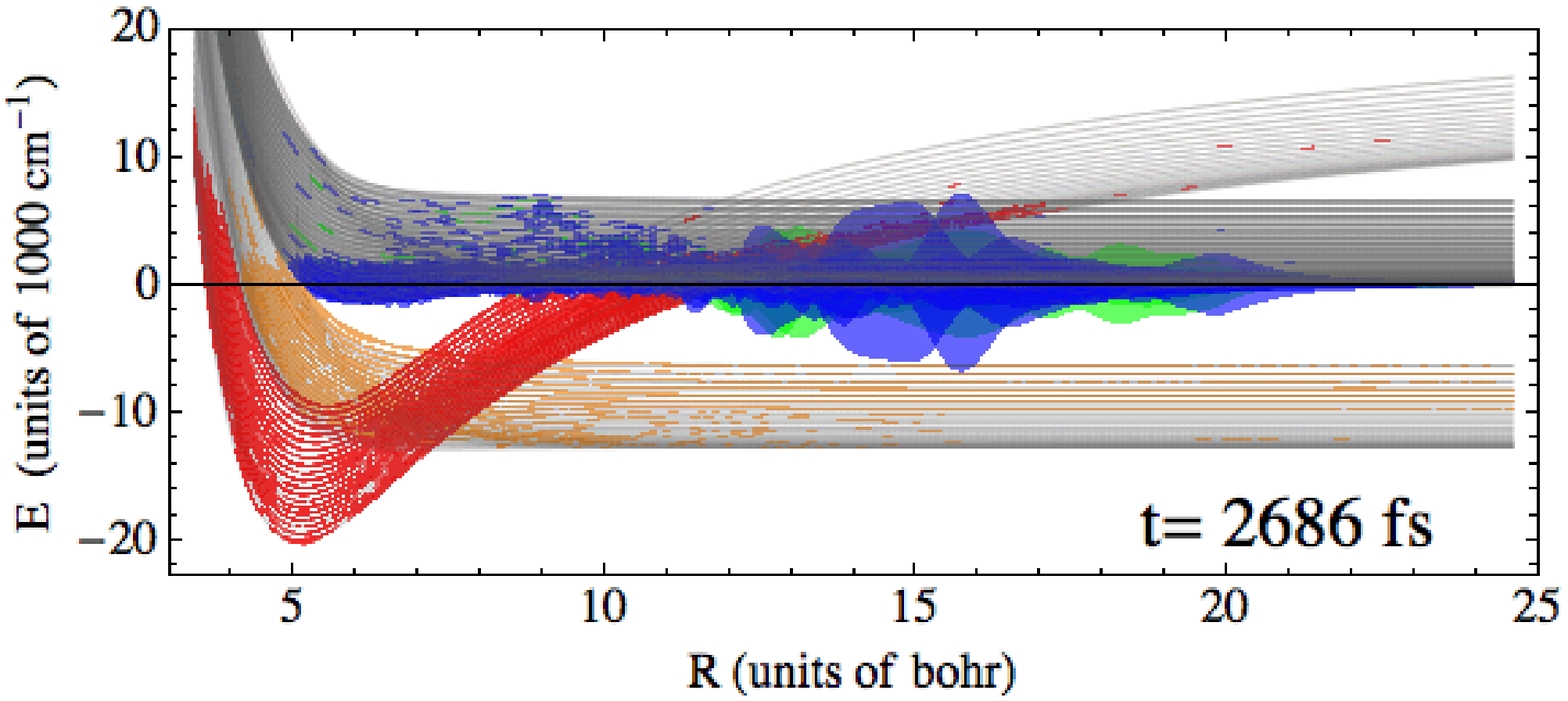}} &
\resizebox{1.0\columnwidth}{!}{\includegraphics{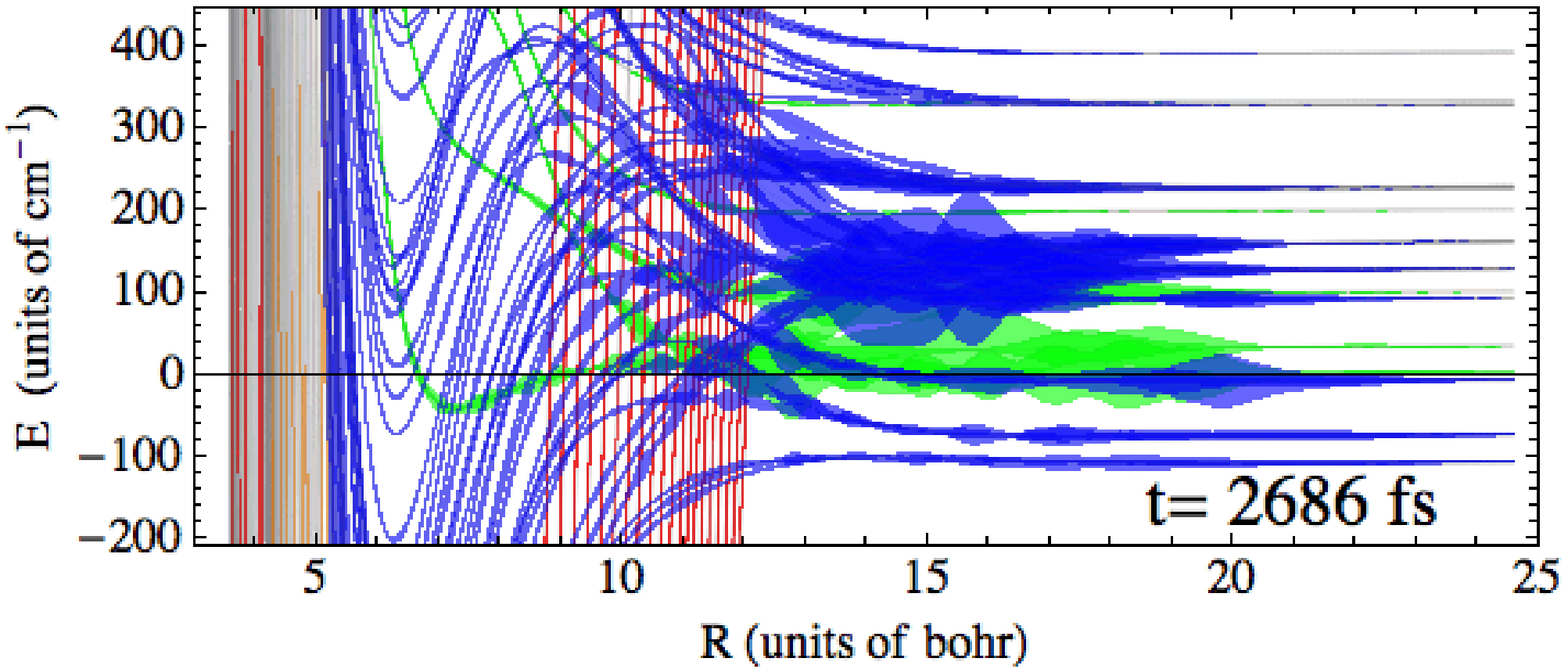}} \\
\resizebox{1.0\columnwidth}{!}{\includegraphics{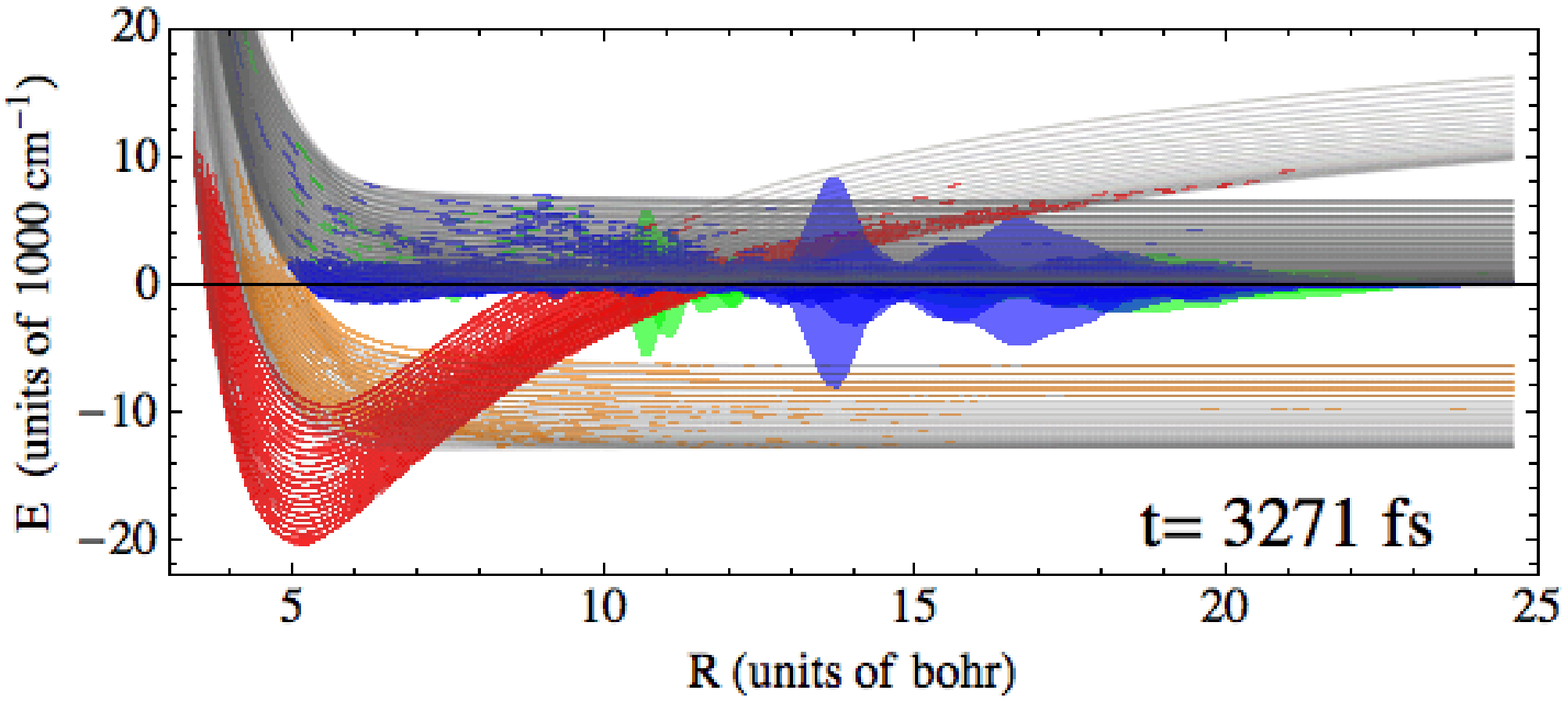}} &
\resizebox{1.0\columnwidth}{!}{\includegraphics{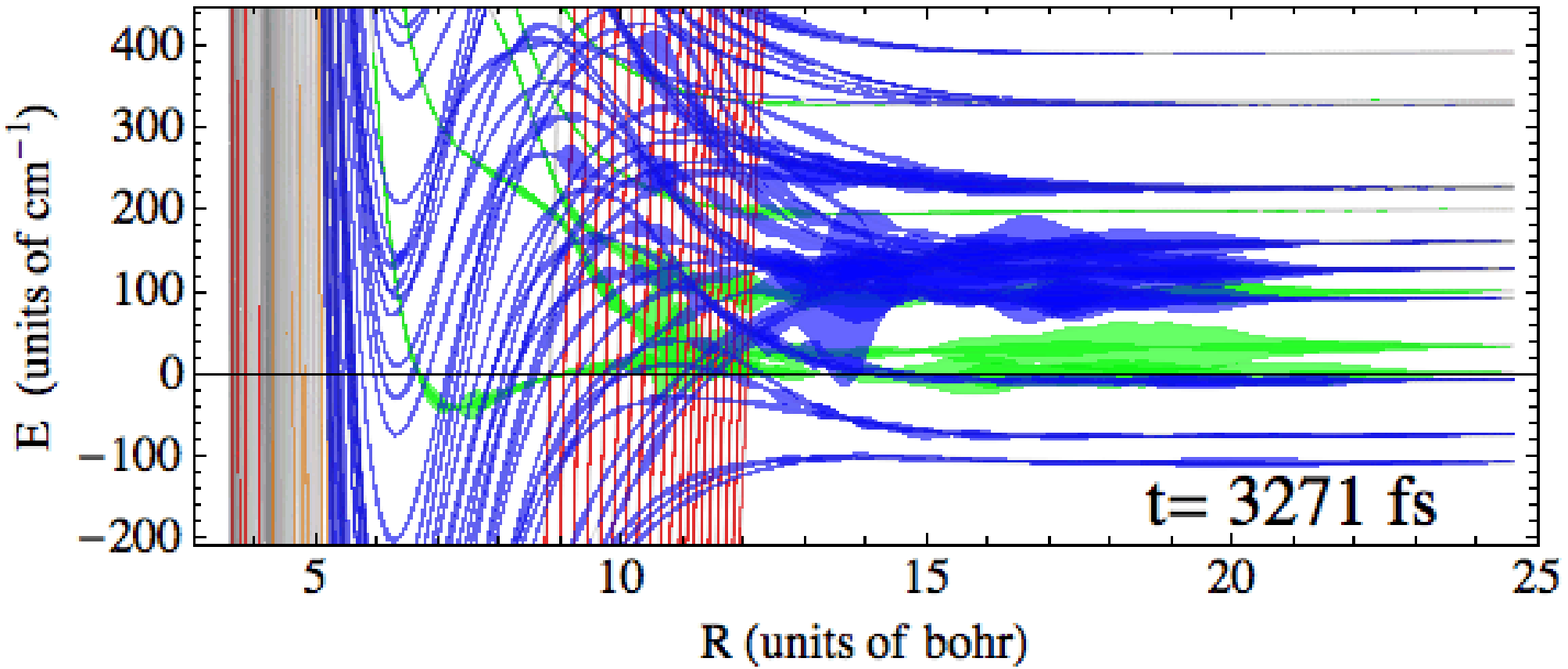}} 
\end{tabular}
\end{center}
\caption{Time-dependent wavepacket propagation described in the text.  
Each frame is a snapshot
of the wavepackets on each adiabatic curve.  The curves are drawn with black lines and the wavepackets
on each of these curves is drawn above and below the corresponding curve.  
Each adiabatic curve corresponds to a different
diabatic electronic channel and these are distinguished by the color of the wavepacket (online only).  Green denotes the $^2\Delta$ channels including the incident one.  Blue denotes Rb ($^2P$) channels; red 
denotes the anion state; and orange denotes the ground state $^2\Sigma$ and $^4\Sigma$ channels.
The wavepacket is incident in
the ground rotational $^2\Delta \ \vert\Omega\vert=3/2$ channel.\label{propfig2}}
\end{figure*}

Time-dependent calculations help to illustrate the dynamics that drive the quenching reaction.  
These have been carried out using a smaller basis of 20 adiabatic basis functions in the Jacobi angle $\gamma$ using 50 
Gauss-Legendre DVR basis functions, due to memory constraints, compared with 42 and 80 for the converged
R-matrix calculation.  The propagated incident gaussian wavepacket, with 
a width of $\sqrt{2}$ bohr in the $R$ degree of freedom, starts in the lowest adiabatic (in $R$) channel of the $^2\Delta$ 
$\vert\Omega\vert=\frac{3}{2}$ electronic state, 
at a radius of $R$=22$a_0$.  The wavepacket is given
an initial translational energy of 90 cm$^{-1}$, such that the lower Rb $^2$P$_{1/2}$ channels are open, but
the $^2$P$_{3/2}$ channels are closed, though the bandwidth of the initial wavepacket is 18 cm$^{-1}$.  We employ a Lanczos propagator of order 12.

The results are shown in Figs.~\ref{propfig} and~\ref{propfig2}.  Fig.~\ref{propfig} shows the probability density
as a function of time and the scattering coordinate $R$, summed over the adiabatic states (in $R$)
for each diabatic electronic channel.
In this figure, the initial state corresponds to the upper left hand panel.
In Fig.~\ref{propfig2}, the initial state is colored green (online).
Both plots show that the density on the initial state electronic surface does not penetrate much
beyond $R$=10.0$a_0$; instead, it couples to several of the Rb $^2$P surfaces and also reflects.  
The cut of the surfaces at linear geometry 
in Fig.~\ref{finalfig}b shows the innermost extent of the attractive part of the initial state $^2\Delta$ surface, near $R$=7.0$a_0$,
whereas for the opposite Rb-H-N geometry the potential well extends only to approximately 10.0$a_0$.  It 
is therefore 
likely that the reflected part of the initial wavepacket, most clearly visible in Fig.~\ref{propfig}, 
comes from configurations near linear Rb-H-N
geometry, and that the remaining flux is lost to the other surfaces within 10.0$a_0$.

Some immediate coupling evidently occurs among
the initial state and all of the Rb ($^2P$) states, as is clear from the similar shape of the 
lowest contour line
in the corrseponding panels of Fig.~\ref{propfig} around 1000 fs.  The coupling seems to be
strongest between the initial $^2\Delta$ state and the Rb ($^2$P$_{1/2}$) $\times$ NH ($^3\Sigma^-$ $\Sigma^{NH}$ = 0), $\vert\Omega\vert=\frac{1}{2}$ state, the eighth state in Table~\ref{so_diatable} and Fig.~\ref{propfig}.  The transition
between the initial Rb ($^2$S) $\times$ NH ($^1\Delta$) and this one is the only transition corresponding to a conservation of the projection of
spin angular momentum on the NH, $\Sigma^{NH}$, along with a transition to the Rb $^2$P$_{1/2}$ state.
This electronic transition does not conserve
the projection of the electronic angular momentum on the NH axis 
and therefore must be driven at nonlinear geometries.

Fig.~\ref{propfig2} shows the density on each adiabatic (in $R$) curve, colored (online) according to the 
electronic channel index.  In viewing this figure one should keep in mind that there is electronic coupling
between these curves, which represent the energies of adiabatic (in $R$) basis functions calculated
on each diabatic electronic surface.  In particular, the coupling among some of the excited state curves and 
the ground state sigma curves means that there is repulsion on the excited state curves and attraction on
the ground state curves that is not represented in this figure.  

This figure shows that the coupling is indeed strongest, earliest, to one of the Rb ($^2P_{1/2}$) 
channels, that which rises, from large $R$ to small $R$, around 12$a_0$, and that correlates to
Rb ($^2P_{1/2}$) + NH ($^3\Sigma^- \ j=1$).  Although some red and orange (online) are visible on the left hand side
of this figure, showing the large view including all of the corresponding 
anion and ground state sigma curves, the bulk of the density resides clearly on electronic states
involved in the quenching reaction.

Despite the fact that the wavefunction amplitude on the anion $^2\Pi$ curves never achieves a large value,
these states have significant influence upon the dynamics, as would be expected from the
large avoided crossings that they create with some of the Rb $^2$P surfaces in the electronically
adiabatic picture.  R-matrix calculations performed
without these diabatic states produced electronically inelastic 
cross sections markedly different (and smaller) than the ones 
shown in Fig.~\ref{scatfig}.  The coupling to the anion surfaces clearly involves a large amount of 
rotational excitation of the NH fragment, as can be seen from the fact that all of the adiabatic curves
corresponding to this electronic state have amplitude on them, colored red (online) in Fig.~\ref{propfig2}.

Because the time-dependent calculation shows that the transition is strongest from the initial state to a 
Rb $^2$P$_{1/2}$ state,
it corroborates our estimate that the avoided crossings around 14.7$a_0$ between components of the
higher-energy, closed Rb $^2$P$_{3/2}$ surfaces and the $^2\Delta$ states, caused by the long-range
dipole-quadrupole interaction, are sufficiently sharp such that they do not provide a good coupling mechanism
for the quenching reaction (at least for the energy considered in this 
time-dependent calculation).  Instead, the evidence suggests that the avoided
crossings and conical intersections involving the $^2\Delta$ 
surface and the Rb $^2$P surfaces further in, caused by
the fact that the anion surface drops down steeply and undergoes broad avoided crossings with these
surfaces, drive the wavepacket onto the the Rb $^2$P$_{1/2}$ surfaces. 
A more thorough analysis is deferred to a forthcoming publication.


The time dependent treatment corroborates the result of the R-matrix calculations, that 
the coupling to the ground NH ($^3\Sigma^-$) $\times$ Rb ($^2$S) electronic state is relatively
small, whereby most of the outgoing flux avoids these channels. The quartet components of this state
seem uninvolved in the quenching reaction for the energy studied, as is clear from
Fig.~\ref{propfig}.  
A modest flux is seen in the doublet ground state sigma component, much less than in the 
Rb $^2$P channels, in agreement with the results of the R-matrix 
calculations.
Thus, it appears that the quenching reaction is both efficient and selective.

\section{Conclusion}

These preliminary results show that the quenching reaction
Rb ($^2$S) + NH ( $^1\Delta$ ) $\rightarrow$ Rb ($^2$P$_{1/2}$) + NH ($X ^3\Sigma^-$) does indeed proceed with high probability, as may have been
expected
due to the near degeneracy of the electronic states involved.  The reaction is
selective, yielding markedly less ground electronic state products.  The coupling
mechanism is seen to involve the complicated set of avoided crossings and conical 
intersections that occur inwards of 14.7$a_0$, not those due to the long-range interactions.
A more complete
study will be published in the future, thoroughly analyzing the dynamics
that occurs between the many electronic states involved in the system, analyzing
the effect of the mixing of electronic and spin angular momentum with 
rotational angular momentum (the Renner-Teller and spin Renner-Teller effects), and
providing reaction rates for comparison with experiment.

\section*{Acknowledgments}
 
We thank Edmund Meyer and V. Kokoouline for helpful discussions.
DJH, SAW and CHG acknowledge support by the DOE Office of Science and by NSF 
grant numbers ITR 0427376, PHY 0427460 and PHY 0427376.
CHG acknowledges additional support from the Alexander von Humboldt Foundation.
HJL acknowledges support under grant AFOSR of the Petroleum Research Fund and NSF grant number PHY 0748742.

\bibliography{RbNH.surf.bib}

\begin{thebibliography}{59}
\expandafter\ifx\csname natexlab\endcsname\relax\def\natexlab#1{#1}\fi
\expandafter\ifx\csname bibnamefont\endcsname\relax
  \def\bibnamefont#1{#1}\fi
\expandafter\ifx\csname bibfnamefont\endcsname\relax
  \def\bibfnamefont#1{#1}\fi
\expandafter\ifx\csname citenamefont\endcsname\relax
  \def\citenamefont#1{#1}\fi
\expandafter\ifx\csname url\endcsname\relax
  \def\url#1{\texttt{#1}}\fi
\expandafter\ifx\csname urlprefix\endcsname\relax\def\urlprefix{URL }\fi
\providecommand{\bibinfo}[2]{#2}
\providecommand{\eprint}[2][]{\url{#2}}

\bibitem[{\citenamefont{Hack}(1998)}]{heather1}
\bibinfo{author}{\bibfnamefont{W.}~\bibnamefont{Hack}}, in
  \emph{\bibinfo{booktitle}{N-Centered Radicals}}, edited by
  \bibinfo{editor}{\bibfnamefont{Z.~B.} \bibnamefont{Alfassi}}
  (\bibinfo{publisher}{Wiley}, \bibinfo{address}{Chirchester},
  \bibinfo{year}{1998}), pp. \bibinfo{pages}{413--466}.

\bibitem[{\citenamefont{Mann and Williams}(1984)}]{heather2}
\bibinfo{author}{\bibfnamefont{A.~P.~C.} \bibnamefont{Mann}} \bibnamefont{and}
  \bibinfo{author}{\bibfnamefont{D.~A.} \bibnamefont{Williams}},
  \bibinfo{journal}{Monthy Notices of the Royal Astronomical Society}
  \textbf{\bibinfo{volume}{209}}, \bibinfo{pages}{33} (\bibinfo{year}{1984}).

\bibitem[{\citenamefont{Smith}(2006)}]{heather4}
\bibinfo{author}{\bibfnamefont{I.~W.~M.} \bibnamefont{Smith}},
  \bibinfo{journal}{Angewandie Chemie-International Edition}
  \textbf{\bibinfo{volume}{45}}, \bibinfo{pages}{2842} (\bibinfo{year}{2006}).

\bibitem[{\citenamefont{Anderson et~al.}(1982)\citenamefont{Anderson, Decker,
  and Kotlar}}]{heather5}
\bibinfo{author}{\bibfnamefont{W.~R.} \bibnamefont{Anderson}},
  \bibinfo{author}{\bibfnamefont{L.~J.} \bibnamefont{Decker}},
  \bibnamefont{and} \bibinfo{author}{\bibfnamefont{A.~J.}
  \bibnamefont{Kotlar}}, \bibinfo{journal}{Combustion and Flame}
  \textbf{\bibinfo{volume}{48}}, \bibinfo{pages}{179} (\bibinfo{year}{1982}).

\bibitem[{\citenamefont{Rinnenthal and Gericke}(1999{\natexlab{a}})}]{heather6}
\bibinfo{author}{\bibfnamefont{J.~L.} \bibnamefont{Rinnenthal}}
  \bibnamefont{and} \bibinfo{author}{\bibfnamefont{K.~H.}
  \bibnamefont{Gericke}}, \bibinfo{journal}{J. Mol. Spectroscopy}
  \textbf{\bibinfo{volume}{198}}, \bibinfo{pages}{115}
  (\bibinfo{year}{1999}{\natexlab{a}}).

\bibitem[{\citenamefont{Slanger and Copeland}(2003)}]{heather7}
\bibinfo{author}{\bibfnamefont{T.~G.} \bibnamefont{Slanger}} \bibnamefont{and}
  \bibinfo{author}{\bibfnamefont{R.~A.} \bibnamefont{Copeland}},
  \bibinfo{journal}{Chem. Rev.} \textbf{\bibinfo{volume}{103}},
  \bibinfo{pages}{4731} (\bibinfo{year}{2003}).

\bibitem[{\citenamefont{Balakrishnan and Dalgarno}(2001)}]{chemultra}
\bibinfo{author}{\bibfnamefont{N.}~\bibnamefont{Balakrishnan}}
  \bibnamefont{and} \bibinfo{author}{\bibfnamefont{A.}~\bibnamefont{Dalgarno}},
  \bibinfo{journal}{Chem. Phys. Lett.} \textbf{\bibinfo{volume}{341}},
  \bibinfo{pages}{652} (\bibinfo{year}{2001}).

\bibitem[{\citenamefont{Burnett et~al.}(2002)\citenamefont{Burnett, Julienne,
  Lett, Tiesinga, and Williams}}]{coldkind}
\bibinfo{author}{\bibfnamefont{K.}~\bibnamefont{Burnett}},
  \bibinfo{author}{\bibfnamefont{P.~S.} \bibnamefont{Julienne}},
  \bibinfo{author}{\bibfnamefont{P.~D.} \bibnamefont{Lett}},
  \bibinfo{author}{\bibfnamefont{E.}~\bibnamefont{Tiesinga}}, \bibnamefont{and}
  \bibinfo{author}{\bibfnamefont{C.~J.} \bibnamefont{Williams}},
  \bibinfo{journal}{Nature (London)} \textbf{\bibinfo{volume}{416}},
  \bibinfo{pages}{225} (\bibinfo{year}{2002}).

\bibitem[{\citenamefont{Zoller}(2002)}]{makingit}
\bibinfo{author}{\bibfnamefont{P.}~\bibnamefont{Zoller}},
  \bibinfo{journal}{Nature (London)} \textbf{\bibinfo{volume}{417}},
  \bibinfo{pages}{493} (\bibinfo{year}{2002}).

\bibitem[{\citenamefont{Zwierlein et~al.}(2003)\citenamefont{Zwierlein, Stan,
  Schunck, Raupach, Gupta, Hadzibabic, and Ketterle}}]{BEmol}
\bibinfo{author}{\bibfnamefont{M.~W.} \bibnamefont{Zwierlein}},
  \bibinfo{author}{\bibfnamefont{C.~A.} \bibnamefont{Stan}},
  \bibinfo{author}{\bibfnamefont{C.~H.} \bibnamefont{Schunck}},
  \bibinfo{author}{\bibfnamefont{S.~M.~F.} \bibnamefont{Raupach}},
  \bibinfo{author}{\bibfnamefont{S.}~\bibnamefont{Gupta}},
  \bibinfo{author}{\bibfnamefont{Z.}~\bibnamefont{Hadzibabic}},
  \bibnamefont{and} \bibinfo{author}{\bibfnamefont{W.}~\bibnamefont{Ketterle}},
  \bibinfo{journal}{Phys. Rev. Lett.} \textbf{\bibinfo{volume}{91}},
  \bibinfo{pages}{250401} (\bibinfo{year}{2003}).

\bibitem[{\citenamefont{Bochinski et~al.}(2004)\citenamefont{Bochinski, Hudson,
  Lewandowski, and Ye}}]{labframe}
\bibinfo{author}{\bibfnamefont{J.~R.} \bibnamefont{Bochinski}},
  \bibinfo{author}{\bibfnamefont{E.~R.} \bibnamefont{Hudson}},
  \bibinfo{author}{\bibfnamefont{H.~J.} \bibnamefont{Lewandowski}},
  \bibnamefont{and} \bibinfo{author}{\bibfnamefont{J.}~\bibnamefont{Ye}},
  \bibinfo{journal}{Phys. Rev. A} \textbf{\bibinfo{volume}{70}},
  \bibinfo{pages}{043410} (\bibinfo{year}{2004}).

\bibitem[{\citenamefont{Krems}(2005)}]{nearzero}
\bibinfo{author}{\bibfnamefont{R.~V.} \bibnamefont{Krems}},
  \bibinfo{journal}{Int. Rev. Phys. Chem.} \textbf{\bibinfo{volume}{24}},
  \bibinfo{pages}{99} (\bibinfo{year}{2005}).

\bibitem[{\citenamefont{Qui et~al.}(2006)\citenamefont{Qui, Ren, Che, Dai,
  Harich, Wang, Yang, Xu, Xie, Gustaffson et~al.}}]{FH2}
\bibinfo{author}{\bibfnamefont{M.}~\bibnamefont{Qui}},
  \bibinfo{author}{\bibfnamefont{Z.}~\bibnamefont{Ren}},
  \bibinfo{author}{\bibfnamefont{L.}~\bibnamefont{Che}},
  \bibinfo{author}{\bibfnamefont{D.}~\bibnamefont{Dai}},
  \bibinfo{author}{\bibfnamefont{S.~A.} \bibnamefont{Harich}},
  \bibinfo{author}{\bibfnamefont{X.}~\bibnamefont{Wang}},
  \bibinfo{author}{\bibfnamefont{X.}~\bibnamefont{Yang}},
  \bibinfo{author}{\bibfnamefont{C.}~\bibnamefont{Xu}},
  \bibinfo{author}{\bibfnamefont{D.}~\bibnamefont{Xie}},
  \bibinfo{author}{\bibfnamefont{M.}~\bibnamefont{Gustaffson}},
  \bibnamefont{et~al.}, \bibinfo{journal}{Science}
  \textbf{\bibinfo{volume}{311}}, \bibinfo{pages}{1440} (\bibinfo{year}{2006}).

\bibitem[{\citenamefont{Anderson et~al.}(1995)\citenamefont{Anderson, Ensher,
  Matthews, Weiman, and Cornell}}]{BEC}
\bibinfo{author}{\bibfnamefont{M.~H.} \bibnamefont{Anderson}},
  \bibinfo{author}{\bibfnamefont{J.~R.} \bibnamefont{Ensher}},
  \bibinfo{author}{\bibfnamefont{M.~R.} \bibnamefont{Matthews}},
  \bibinfo{author}{\bibfnamefont{C.~E.} \bibnamefont{Weiman}},
  \bibnamefont{and} \bibinfo{author}{\bibfnamefont{E.~A.}
  \bibnamefont{Cornell}}, \bibinfo{journal}{Science}
  \textbf{\bibinfo{volume}{269}}, \bibinfo{pages}{198} (\bibinfo{year}{1995}).

\bibitem[{\citenamefont{Myatt et~al.}(1997)\citenamefont{Myatt, Burt, Ghrist,
  Cornell, and Weiman}}]{overlapping}
\bibinfo{author}{\bibfnamefont{C.~J.} \bibnamefont{Myatt}},
  \bibinfo{author}{\bibfnamefont{E.~A.} \bibnamefont{Burt}},
  \bibinfo{author}{\bibfnamefont{R.~W.} \bibnamefont{Ghrist}},
  \bibinfo{author}{\bibfnamefont{E.~A.} \bibnamefont{Cornell}},
  \bibnamefont{and} \bibinfo{author}{\bibfnamefont{C.~E.}
  \bibnamefont{Weiman}}, \bibinfo{journal}{Phys. Rev. Lett.}
  \textbf{\bibinfo{volume}{78}}, \bibinfo{pages}{586} (\bibinfo{year}{1997}).

\bibitem[{\citenamefont{Weinstein et~al.}(1998)\citenamefont{Weinstein,
  deCarvalho, Guillet, Friedrich, and Doyle}}]{cah}
\bibinfo{author}{\bibfnamefont{J.~D.} \bibnamefont{Weinstein}},
  \bibinfo{author}{\bibfnamefont{R.}~\bibnamefont{deCarvalho}},
  \bibinfo{author}{\bibfnamefont{T.}~\bibnamefont{Guillet}},
  \bibinfo{author}{\bibfnamefont{B.}~\bibnamefont{Friedrich}},
  \bibnamefont{and} \bibinfo{author}{\bibfnamefont{J.~M.} \bibnamefont{Doyle}},
  \bibinfo{journal}{Nature (London)} \textbf{\bibinfo{volume}{395}},
  \bibinfo{pages}{148} (\bibinfo{year}{1998}).

\bibitem[{\citenamefont{Santos et~al.}(2000)\citenamefont{Santos, Shiyapnikov,
  Zoller, and Lewenstein}}]{BECdipole}
\bibinfo{author}{\bibfnamefont{L.}~\bibnamefont{Santos}},
  \bibinfo{author}{\bibfnamefont{G.~V.} \bibnamefont{Shiyapnikov}},
  \bibinfo{author}{\bibfnamefont{P.}~\bibnamefont{Zoller}}, \bibnamefont{and}
  \bibinfo{author}{\bibfnamefont{M.}~\bibnamefont{Lewenstein}},
  \bibinfo{journal}{Phys. Rev. Lett.} \textbf{\bibinfo{volume}{85}},
  \bibinfo{pages}{1791} (\bibinfo{year}{2000}).

\bibitem[{\citenamefont{Wang et~al.}(2004)\citenamefont{Wang, Qi, Stone,
  Nikolayeva, Wang, Hattaway, Gensemer, Gould, Eyler, and Stwalley}}]{KRb}
\bibinfo{author}{\bibfnamefont{D.}~\bibnamefont{Wang}},
  \bibinfo{author}{\bibfnamefont{J.}~\bibnamefont{Qi}},
  \bibinfo{author}{\bibfnamefont{M.~F.} \bibnamefont{Stone}},
  \bibinfo{author}{\bibfnamefont{O.}~\bibnamefont{Nikolayeva}},
  \bibinfo{author}{\bibfnamefont{H.}~\bibnamefont{Wang}},
  \bibinfo{author}{\bibfnamefont{B.}~\bibnamefont{Hattaway}},
  \bibinfo{author}{\bibfnamefont{S.~D.} \bibnamefont{Gensemer}},
  \bibinfo{author}{\bibfnamefont{P.~L.} \bibnamefont{Gould}},
  \bibinfo{author}{\bibfnamefont{E.~E.} \bibnamefont{Eyler}}, \bibnamefont{and}
  \bibinfo{author}{\bibfnamefont{W.~C.} \bibnamefont{Stwalley}},
  \bibinfo{journal}{Phys. Rev. Lett.} \textbf{\bibinfo{volume}{93}},
  \bibinfo{pages}{243005} (\bibinfo{year}{2004}).

\bibitem[{\citenamefont{Sage et~al.}(2005)\citenamefont{Sage, Sainis, Bergeman,
  and DeMille}}]{opticalmol}
\bibinfo{author}{\bibfnamefont{J.~M.} \bibnamefont{Sage}},
  \bibinfo{author}{\bibfnamefont{S.}~\bibnamefont{Sainis}},
  \bibinfo{author}{\bibfnamefont{T.}~\bibnamefont{Bergeman}}, \bibnamefont{and}
  \bibinfo{author}{\bibfnamefont{D.}~\bibnamefont{DeMille}},
  \bibinfo{journal}{Phys. Rev. Lett.} \textbf{\bibinfo{volume}{94}},
  \bibinfo{pages}{203001} (\bibinfo{year}{2005}).

\bibitem[{\citenamefont{Sawyer et~al.}(2007)\citenamefont{Sawyer, Lev, Hudson,
  Stuhl, Lara, Bohn, and Ye}}]{trapping_oh}
\bibinfo{author}{\bibfnamefont{B.~C.} \bibnamefont{Sawyer}},
  \bibinfo{author}{\bibfnamefont{B.~L.} \bibnamefont{Lev}},
  \bibinfo{author}{\bibfnamefont{E.~R.} \bibnamefont{Hudson}},
  \bibinfo{author}{\bibfnamefont{B.~K.} \bibnamefont{Stuhl}},
  \bibinfo{author}{\bibfnamefont{M.}~\bibnamefont{Lara}},
  \bibinfo{author}{\bibfnamefont{J.~L.} \bibnamefont{Bohn}}, \bibnamefont{and}
  \bibinfo{author}{\bibfnamefont{J.}~\bibnamefont{Ye}}, \bibinfo{journal}{Phys.
  Rev. Lett.} \textbf{\bibinfo{volume}{98}}, \bibinfo{pages}{253002}
  (\bibinfo{year}{2007}).

\bibitem[{\citenamefont{van~de Meerakker et~al.}(2001)\citenamefont{van~de
  Meerakker, Jongma, Bethlem, and Meijer}}]{magNH2}
\bibinfo{author}{\bibfnamefont{S.~Y.~T.} \bibnamefont{van~de Meerakker}},
  \bibinfo{author}{\bibfnamefont{R.~T.} \bibnamefont{Jongma}},
  \bibinfo{author}{\bibfnamefont{H.~L.} \bibnamefont{Bethlem}},
  \bibnamefont{and} \bibinfo{author}{\bibfnamefont{G.}~\bibnamefont{Meijer}},
  \bibinfo{journal}{Phys. Rev. A} \textbf{\bibinfo{volume}{64}},
  \bibinfo{pages}{041401} (\bibinfo{year}{2001}).

\bibitem[{\citenamefont{Egorov et~al.}(2004)\citenamefont{Egorov, Campbell,
  Friedrich, Maxwell, Tsikata, van Buuren, and Doyle}}]{bufferNH}
\bibinfo{author}{\bibfnamefont{D.}~\bibnamefont{Egorov}},
  \bibinfo{author}{\bibfnamefont{W.~C.} \bibnamefont{Campbell}},
  \bibinfo{author}{\bibfnamefont{B.}~\bibnamefont{Friedrich}},
  \bibinfo{author}{\bibfnamefont{S.~E.} \bibnamefont{Maxwell}},
  \bibinfo{author}{\bibfnamefont{E.}~\bibnamefont{Tsikata}},
  \bibinfo{author}{\bibfnamefont{L.~D.} \bibnamefont{van Buuren}},
  \bibnamefont{and} \bibinfo{author}{\bibfnamefont{J.~M.} \bibnamefont{Doyle}},
  \bibinfo{journal}{Eur. Phys. J. D} \textbf{\bibinfo{volume}{31}},
  \bibinfo{pages}{307} (\bibinfo{year}{2004}).

\bibitem[{\citenamefont{Campbell et~al.}(2007)\citenamefont{Campbell, Tsikata,
  Lu, van Buuren, and Doyle}}]{magNH}
\bibinfo{author}{\bibfnamefont{W.~C.} \bibnamefont{Campbell}},
  \bibinfo{author}{\bibfnamefont{E.}~\bibnamefont{Tsikata}},
  \bibinfo{author}{\bibfnamefont{H.-I.} \bibnamefont{Lu}},
  \bibinfo{author}{\bibfnamefont{L.~D.} \bibnamefont{van Buuren}},
  \bibnamefont{and} \bibinfo{author}{\bibfnamefont{J.~M.} \bibnamefont{Doyle}},
  \bibinfo{journal}{Phys. Rev. Lett.} \textbf{\bibinfo{volume}{98}},
  \bibinfo{pages}{213001} (\bibinfo{year}{2007}).

\bibitem[{\citenamefont{Hoekstra et~al.}(2007)\citenamefont{Hoekstra, Metsala,
  Zieger, Sharfenberg, Gilijamse, Meijer, and van~de Meerakker}}]{trapping_nh}
\bibinfo{author}{\bibfnamefont{S.}~\bibnamefont{Hoekstra}},
  \bibinfo{author}{\bibfnamefont{M.}~\bibnamefont{Metsala}},
  \bibinfo{author}{\bibfnamefont{P.~C.} \bibnamefont{Zieger}},
  \bibinfo{author}{\bibfnamefont{L.}~\bibnamefont{Sharfenberg}},
  \bibinfo{author}{\bibfnamefont{J.~J.} \bibnamefont{Gilijamse}},
  \bibinfo{author}{\bibfnamefont{G.}~\bibnamefont{Meijer}}, \bibnamefont{and}
  \bibinfo{author}{\bibfnamefont{S.~Y.~T.} \bibnamefont{van~de Meerakker}},
  \bibinfo{journal}{Phys. Rev. A} \textbf{\bibinfo{volume}{76}},
  \bibinfo{pages}{063408} (\bibinfo{year}{2007}).

\bibitem[{\citenamefont{Hummon et~al.}(2008)\citenamefont{Hummon, Campbell, Lu,
  Tsikata, Wang, and Doyle}}]{cotrapping_n_nh}
\bibinfo{author}{\bibfnamefont{M.~T.} \bibnamefont{Hummon}},
  \bibinfo{author}{\bibfnamefont{W.~C.} \bibnamefont{Campbell}},
  \bibinfo{author}{\bibfnamefont{H.-I.} \bibnamefont{Lu}},
  \bibinfo{author}{\bibfnamefont{E.}~\bibnamefont{Tsikata}},
  \bibinfo{author}{\bibfnamefont{Y.}~\bibnamefont{Wang}}, \bibnamefont{and}
  \bibinfo{author}{\bibfnamefont{J.~M.} \bibnamefont{Doyle}},
  \bibinfo{journal}{Phys. Rev. A} \textbf{\bibinfo{volume}{78}},
  \bibinfo{pages}{050702} (\bibinfo{year}{2008}).

\bibitem[{\citenamefont{P.~Soldan and Hutson}(2009)}]{hutson_hope}
\bibinfo{author}{\bibfnamefont{P.~S.~Z.} \bibnamefont{P.~Soldan}}
  \bibnamefont{and} \bibinfo{author}{\bibfnamefont{J.~M.} \bibnamefont{Hutson}}
  (\bibinfo{year}{2009}), \eprint{Arxiv:0901.2493}.

\bibitem[{\citenamefont{Soldan and Hutson}(2004)}]{hutson_rbnh}
\bibinfo{author}{\bibfnamefont{P.}~\bibnamefont{Soldan}} \bibnamefont{and}
  \bibinfo{author}{\bibfnamefont{J.~M.} \bibnamefont{Hutson}},
  \bibinfo{journal}{Phys. Rev. Lett.} \textbf{\bibinfo{volume}{92}},
  \bibinfo{pages}{163202} (\bibinfo{year}{2004}).

\bibitem[{\citenamefont{Tacconi
  et~al.}(2007{\natexlab{a}})\citenamefont{Tacconi, Gonzalez-Sanchez, Bodo, and
  Gianturco}}]{gianturco_ultralow}
\bibinfo{author}{\bibfnamefont{M.}~\bibnamefont{Tacconi}},
  \bibinfo{author}{\bibfnamefont{L.}~\bibnamefont{Gonzalez-Sanchez}},
  \bibinfo{author}{\bibfnamefont{E.}~\bibnamefont{Bodo}}, \bibnamefont{and}
  \bibinfo{author}{\bibfnamefont{F.~A.} \bibnamefont{Gianturco}},
  \bibinfo{journal}{Phys. Rev. A} \textbf{\bibinfo{volume}{76}},
  \bibinfo{pages}{032702} (\bibinfo{year}{2007}{\natexlab{a}}).

\bibitem[{\citenamefont{Tacconi
  et~al.}(2007{\natexlab{b}})\citenamefont{Tacconi, Bodo, and
  Gianturco}}]{gianturco_accounts}
\bibinfo{author}{\bibfnamefont{M.}~\bibnamefont{Tacconi}},
  \bibinfo{author}{\bibfnamefont{E.}~\bibnamefont{Bodo}}, \bibnamefont{and}
  \bibinfo{author}{\bibfnamefont{F.~A.} \bibnamefont{Gianturco}},
  \bibinfo{journal}{Theor. Chem. Acc.} \textbf{\bibinfo{volume}{117}},
  \bibinfo{pages}{649} (\bibinfo{year}{2007}{\natexlab{b}}).

\bibitem[{\citenamefont{Lara et~al.}(2006)\citenamefont{Lara, Bohn, Potter,
  Soldan, and Hutson}}]{hutson_rboh_prospects}
\bibinfo{author}{\bibfnamefont{M.}~\bibnamefont{Lara}},
  \bibinfo{author}{\bibfnamefont{J.~L.} \bibnamefont{Bohn}},
  \bibinfo{author}{\bibfnamefont{D.}~\bibnamefont{Potter}},
  \bibinfo{author}{\bibfnamefont{P.}~\bibnamefont{Soldan}}, \bibnamefont{and}
  \bibinfo{author}{\bibfnamefont{J.~M.} \bibnamefont{Hutson}},
  \bibinfo{journal}{Phys. Rev. Lett.} \textbf{\bibinfo{volume}{97}},
  \bibinfo{pages}{183201} (\bibinfo{year}{2006}).

\bibitem[{\citenamefont{Lara et~al.}(2007)\citenamefont{Lara, Bohn, Potter,
  Soldan, and Hutson}}]{hutson_rboh}
\bibinfo{author}{\bibfnamefont{M.}~\bibnamefont{Lara}},
  \bibinfo{author}{\bibfnamefont{J.~L.} \bibnamefont{Bohn}},
  \bibinfo{author}{\bibfnamefont{D.~E.} \bibnamefont{Potter}},
  \bibinfo{author}{\bibfnamefont{P.}~\bibnamefont{Soldan}}, \bibnamefont{and}
  \bibinfo{author}{\bibfnamefont{J.~M.} \bibnamefont{Hutson}},
  \bibinfo{journal}{Phys. Rev. A} \textbf{\bibinfo{volume}{75}},
  \bibinfo{pages}{012704} (\bibinfo{year}{2007}).

\bibitem[{\citenamefont{Krems et~al.}(2003)\citenamefont{Krems, Sadaghpour,
  Dalgarno, Zgid, Klos, and Chatasinski}}]{HeNHab}
\bibinfo{author}{\bibfnamefont{R.~V.} \bibnamefont{Krems}},
  \bibinfo{author}{\bibfnamefont{H.~R.} \bibnamefont{Sadaghpour}},
  \bibinfo{author}{\bibfnamefont{A.}~\bibnamefont{Dalgarno}},
  \bibinfo{author}{\bibfnamefont{D.}~\bibnamefont{Zgid}},
  \bibinfo{author}{\bibfnamefont{J.}~\bibnamefont{Klos}}, \bibnamefont{and}
  \bibinfo{author}{\bibfnamefont{G.}~\bibnamefont{Chatasinski}},
  \bibinfo{journal}{Phys. Rev. A} \textbf{\bibinfo{volume}{68}},
  \bibinfo{pages}{051401} (\bibinfo{year}{2003}).

\bibitem[{\citenamefont{Cybulski et~al.}(2005)\citenamefont{Cybulski, Krems,
  Sadeghpour, Dalgarno, Kross, Groenenboom, van~der Avoird, Zgid, and
  Chatasinski}}]{HeNH}
\bibinfo{author}{\bibfnamefont{H.}~\bibnamefont{Cybulski}},
  \bibinfo{author}{\bibfnamefont{R.~V.} \bibnamefont{Krems}},
  \bibinfo{author}{\bibfnamefont{H.~R.} \bibnamefont{Sadeghpour}},
  \bibinfo{author}{\bibfnamefont{A.}~\bibnamefont{Dalgarno}},
  \bibinfo{author}{\bibfnamefont{J.}~\bibnamefont{Kross}},
  \bibinfo{author}{\bibfnamefont{G.~C.} \bibnamefont{Groenenboom}},
  \bibinfo{author}{\bibfnamefont{A.}~\bibnamefont{van~der Avoird}},
  \bibinfo{author}{\bibfnamefont{D.}~\bibnamefont{Zgid}}, \bibnamefont{and}
  \bibinfo{author}{\bibfnamefont{G.}~\bibnamefont{Chatasinski}},
  \bibinfo{journal}{J. Chem. Phys.} \textbf{\bibinfo{volume}{122}},
  \bibinfo{pages}{094307} (\bibinfo{year}{2005}).

\bibitem[{\citenamefont{Gonzalez-Martinez and Hutson}(2007)}]{hutson_henh}
\bibinfo{author}{\bibfnamefont{M.~L.} \bibnamefont{Gonzalez-Martinez}}
  \bibnamefont{and} \bibinfo{author}{\bibfnamefont{J.~M.}
  \bibnamefont{Hutson}}, \bibinfo{journal}{Phys. Rev. A}
  \textbf{\bibinfo{volume}{75}}, \bibinfo{pages}{022702}
  (\bibinfo{year}{2007}).

\bibitem[{\citenamefont{Zuchowski and Hutson}(2008)}]{hutson_nh3}
\bibinfo{author}{\bibfnamefont{P.~S.} \bibnamefont{Zuchowski}}
  \bibnamefont{and} \bibinfo{author}{\bibfnamefont{J.~M.}
  \bibnamefont{Hutson}}, \bibinfo{journal}{Phys. Rev. A}
  \textbf{\bibinfo{volume}{78}}, \bibinfo{pages}{022701}
  (\bibinfo{year}{2008}).

\bibitem[{\citenamefont{Zuchowski and Hutson}(2009)}]{hutson_nh3_xxx}
\bibinfo{author}{\bibfnamefont{P.~S.} \bibnamefont{Zuchowski}}
  \bibnamefont{and} \bibinfo{author}{\bibfnamefont{J.~M.} \bibnamefont{Hutson}}
  (\bibinfo{year}{2009}), \eprint{Arxiv:0902.4548}.

\bibitem[{\citenamefont{Dhont et~al.}(2005)\citenamefont{Dhont, van Lenthe,
  Groenenboom, and van~der Avoird}}]{NHNH}
\bibinfo{author}{\bibfnamefont{G.~S.~F.} \bibnamefont{Dhont}},
  \bibinfo{author}{\bibfnamefont{J.~H.} \bibnamefont{van Lenthe}},
  \bibinfo{author}{\bibfnamefont{G.~C.} \bibnamefont{Groenenboom}},
  \bibnamefont{and} \bibinfo{author}{\bibfnamefont{A.}~\bibnamefont{van~der
  Avoird}}, \bibinfo{journal}{J. Chem. Phys.} \textbf{\bibinfo{volume}{123}},
  \bibinfo{pages}{184302} (\bibinfo{year}{2005}).

\bibitem[{\citenamefont{Rinnenthal and Gericke}(1999{\natexlab{b}})}]{data2}
\bibinfo{author}{\bibfnamefont{J.~L.} \bibnamefont{Rinnenthal}}
  \bibnamefont{and} \bibinfo{author}{\bibfnamefont{K.}~\bibnamefont{Gericke}},
  \bibinfo{journal}{J. Mol. Spec.} \textbf{\bibinfo{volume}{198}},
  \bibinfo{pages}{115} (\bibinfo{year}{1999}{\natexlab{b}}).

\bibitem[{\citenamefont{Sansonetti}(2006)}]{data1}
\bibinfo{author}{\bibfnamefont{J.~E.} \bibnamefont{Sansonetti}},
  \bibinfo{journal}{J. Phys. Chem. Ref. Data} \textbf{\bibinfo{volume}{35}},
  \bibinfo{pages}{301} (\bibinfo{year}{2006}).

\bibitem[{\citenamefont{Bethlem et~al.}(1999)\citenamefont{Bethlem, Berden, and
  Meijer}}]{heather8}
\bibinfo{author}{\bibfnamefont{H.~L.} \bibnamefont{Bethlem}},
  \bibinfo{author}{\bibfnamefont{G.}~\bibnamefont{Berden}}, \bibnamefont{and}
  \bibinfo{author}{\bibfnamefont{G.}~\bibnamefont{Meijer}},
  \bibinfo{journal}{Phys. Rev. Lett.} \textbf{\bibinfo{volume}{83}},
  \bibinfo{pages}{1558} (\bibinfo{year}{1999}).

\bibitem[{\citenamefont{van~de Meerakker et~al.}(2005)\citenamefont{van~de
  Meerakker, Smeets, Vanhaecke, Jongma, and Meijer}}]{heather9}
\bibinfo{author}{\bibfnamefont{S.~Y.~T.} \bibnamefont{van~de Meerakker}},
  \bibinfo{author}{\bibfnamefont{P.~H.~M.} \bibnamefont{Smeets}},
  \bibinfo{author}{\bibfnamefont{N.}~\bibnamefont{Vanhaecke}},
  \bibinfo{author}{\bibfnamefont{R.~T.} \bibnamefont{Jongma}},
  \bibnamefont{and} \bibinfo{author}{\bibfnamefont{G.}~\bibnamefont{Meijer}},
  \bibinfo{journal}{Phys. Rev. Lett.} \textbf{\bibinfo{volume}{94}},
  \bibinfo{pages}{023004} (\bibinfo{year}{2005}).

\bibitem[{\citenamefont{Lishka et~al.}(1981)\citenamefont{Lishka, Shepard,
  Brown, and Shavitt}}]{col1}
\bibinfo{author}{\bibfnamefont{H.}~\bibnamefont{Lishka}},
  \bibinfo{author}{\bibfnamefont{R.}~\bibnamefont{Shepard}},
  \bibinfo{author}{\bibfnamefont{F.~B.} \bibnamefont{Brown}}, \bibnamefont{and}
  \bibinfo{author}{\bibfnamefont{I.}~\bibnamefont{Shavitt}},
  \bibinfo{journal}{Int. J. Quantum Chem., Quantum Chem. Symp.}
  \textbf{\bibinfo{volume}{15}}, \bibinfo{pages}{91} (\bibinfo{year}{1981}).

\bibitem[{\citenamefont{Shepard et~al.}(1988)\citenamefont{Shepard, Shavitt,
  Pitzer, Comeau, Pepper, Lishka, Szalay, Ahlrichs, Brown, and Zhao}}]{col2}
\bibinfo{author}{\bibfnamefont{R.}~\bibnamefont{Shepard}},
  \bibinfo{author}{\bibfnamefont{I.}~\bibnamefont{Shavitt}},
  \bibinfo{author}{\bibfnamefont{R.~M.} \bibnamefont{Pitzer}},
  \bibinfo{author}{\bibfnamefont{D.~C.} \bibnamefont{Comeau}},
  \bibinfo{author}{\bibfnamefont{M.}~\bibnamefont{Pepper}},
  \bibinfo{author}{\bibfnamefont{H.}~\bibnamefont{Lishka}},
  \bibinfo{author}{\bibfnamefont{P.~G.} \bibnamefont{Szalay}},
  \bibinfo{author}{\bibfnamefont{R.}~\bibnamefont{Ahlrichs}},
  \bibinfo{author}{\bibfnamefont{F.~B.} \bibnamefont{Brown}}, \bibnamefont{and}
  \bibinfo{author}{\bibfnamefont{J.}~\bibnamefont{Zhao}},
  \bibinfo{journal}{Int. J. Quantum Chem., Quantum Chem. Symp.}
  \textbf{\bibinfo{volume}{22}}, \bibinfo{pages}{149} (\bibinfo{year}{1988}).

\bibitem[{\citenamefont{Lishka et~al.}(2006)\citenamefont{Lishka, Shepard,
  Shavitt, Pitzer, Dallos, Muller, Szalay, Brown, Ahlrichs, Bohm
  et~al.}}]{col3}
\bibinfo{author}{\bibfnamefont{H.}~\bibnamefont{Lishka}},
  \bibinfo{author}{\bibfnamefont{R.}~\bibnamefont{Shepard}},
  \bibinfo{author}{\bibfnamefont{I.}~\bibnamefont{Shavitt}},
  \bibinfo{author}{\bibfnamefont{R.~M.} \bibnamefont{Pitzer}},
  \bibinfo{author}{\bibfnamefont{M.}~\bibnamefont{Dallos}},
  \bibinfo{author}{\bibfnamefont{T.}~\bibnamefont{Muller}},
  \bibinfo{author}{\bibfnamefont{P.~G.} \bibnamefont{Szalay}},
  \bibinfo{author}{\bibfnamefont{F.~B.} \bibnamefont{Brown}},
  \bibinfo{author}{\bibfnamefont{R.}~\bibnamefont{Ahlrichs}},
  \bibinfo{author}{\bibfnamefont{H.~J.} \bibnamefont{Bohm}},
  \bibnamefont{et~al.}, \emph{\bibinfo{title}{Columbus, an ab initio electronic
  structure program}} (\bibinfo{year}{2006}), \bibinfo{note}{release 5.9.1}.

\bibitem[{\citenamefont{v.~Szentpaly et~al.}(1982)\citenamefont{v.~Szentpaly,
  Fuenealba, and Stoll}}]{pseudo1}
\bibinfo{author}{\bibfnamefont{L.}~\bibnamefont{v.~Szentpaly}},
  \bibinfo{author}{\bibfnamefont{P.}~\bibnamefont{Fuenealba}},
  \bibnamefont{and} \bibinfo{author}{\bibfnamefont{H.}~\bibnamefont{Stoll}},
  \bibinfo{journal}{Chem. Phys. Lett.} \textbf{\bibinfo{volume}{93}},
  \bibinfo{pages}{555} (\bibinfo{year}{1982}).

\bibitem[{\citenamefont{Fuentealba et~al.}(1983)\citenamefont{Fuentealba,
  Stoll, v.~Szentpaly, Schwerdtfeger, and Preuss}}]{pseudo2}
\bibinfo{author}{\bibfnamefont{P.}~\bibnamefont{Fuentealba}},
  \bibinfo{author}{\bibfnamefont{H.}~\bibnamefont{Stoll}},
  \bibinfo{author}{\bibfnamefont{L.}~\bibnamefont{v.~Szentpaly}},
  \bibinfo{author}{\bibfnamefont{P.}~\bibnamefont{Schwerdtfeger}},
  \bibnamefont{and} \bibinfo{author}{\bibfnamefont{H.}~\bibnamefont{Preuss}},
  \bibinfo{journal}{J. Phys. B} \textbf{\bibinfo{volume}{16}},
  \bibinfo{pages}{L323} (\bibinfo{year}{1983}).

\bibitem[{\citenamefont{Domcke and Stock}(1997)}]{diareview}
\bibinfo{author}{\bibfnamefont{W.}~\bibnamefont{Domcke}} \bibnamefont{and}
  \bibinfo{author}{\bibfnamefont{G.}~\bibnamefont{Stock}},
  \bibinfo{journal}{Adv. Chem. Phys.} \textbf{\bibinfo{volume}{100}},
  \bibinfo{pages}{1} (\bibinfo{year}{1997}).

\bibitem[{\citenamefont{T.~H.~Dunning}(1989)}]{dunning1}
\bibinfo{author}{\bibfnamefont{J.}~\bibnamefont{T.~H.~Dunning}},
  \bibinfo{journal}{J. Chem. Phys.} \textbf{\bibinfo{volume}{90}},
  \bibinfo{pages}{1007} (\bibinfo{year}{1989}).

\bibitem[{\citenamefont{Macias and Riera}(1978)}]{propertydia}
\bibinfo{author}{\bibfnamefont{A.}~\bibnamefont{Macias}} \bibnamefont{and}
  \bibinfo{author}{\bibfnamefont{A.}~\bibnamefont{Riera}}, \bibinfo{journal}{J.
  Phys. B} \textbf{\bibinfo{volume}{11}}, \bibinfo{pages}{L489}
  (\bibinfo{year}{1978}).

\bibitem[{\citenamefont{Werner and Meyer}(1981)}]{dipoledia}
\bibinfo{author}{\bibfnamefont{H.-J.} \bibnamefont{Werner}} \bibnamefont{and}
  \bibinfo{author}{\bibfnamefont{W.}~\bibnamefont{Meyer}}, \bibinfo{journal}{J.
  Chem. Phys.} \textbf{\bibinfo{volume}{74}}, \bibinfo{pages}{5802}
  (\bibinfo{year}{1981}).

\bibitem[{\citenamefont{Petrongolo}(1988)}]{petrongolo}
\bibinfo{author}{\bibfnamefont{C.}~\bibnamefont{Petrongolo}},
  \bibinfo{journal}{J. Chem. Phys.} \textbf{\bibinfo{volume}{89}},
  \bibinfo{pages}{1297} (\bibinfo{year}{1988}).

\bibitem[{\citenamefont{Sukiasyan and Meyer}(2001)}]{hd2}
\bibinfo{author}{\bibfnamefont{S.}~\bibnamefont{Sukiasyan}} \bibnamefont{and}
  \bibinfo{author}{\bibfnamefont{H.-D.} \bibnamefont{Meyer}},
  \bibinfo{journal}{J. Phys. Chem. A} \textbf{\bibinfo{volume}{105}},
  \bibinfo{pages}{2604} (\bibinfo{year}{2001}).

\bibitem[{\citenamefont{Huber and Herzberg}(1979)}]{herz}
\bibinfo{author}{\bibfnamefont{K.~P.} \bibnamefont{Huber}} \bibnamefont{and}
  \bibinfo{author}{\bibfnamefont{G.}~\bibnamefont{Herzberg}},
  \emph{\bibinfo{title}{Molecular Spectra and Molecular Structure. IV.
  Constants of Diatomic Molecules}} (\bibinfo{publisher}{Van Nostrand Reinhold
  Co.}, \bibinfo{year}{1979}).

\bibitem[{\citenamefont{Baluja et~al.}(1982)\citenamefont{Baluja, Burke, and
  Morgan}}]{propagator}
\bibinfo{author}{\bibfnamefont{K.~L.} \bibnamefont{Baluja}},
  \bibinfo{author}{\bibfnamefont{P.~G.} \bibnamefont{Burke}}, \bibnamefont{and}
  \bibinfo{author}{\bibfnamefont{L.~A.} \bibnamefont{Morgan}},
  \bibinfo{journal}{Computer Physics Communications}
  \textbf{\bibinfo{volume}{27}}, \bibinfo{pages}{299} (\bibinfo{year}{1982}).

\bibitem[{\citenamefont{Dickinson and Certain}(1968)}]{dickcert}
\bibinfo{author}{\bibfnamefont{A.~S.} \bibnamefont{Dickinson}}
  \bibnamefont{and} \bibinfo{author}{\bibfnamefont{P.~R.}
  \bibnamefont{Certain}}, \bibinfo{journal}{J Chem Phys}
  \textbf{\bibinfo{volume}{49}}, \bibinfo{pages}{4209} (\bibinfo{year}{1968}).

\bibitem[{\citenamefont{Light et~al.}(1985)\citenamefont{Light, Hamilton, and
  Lill}}]{lhl}
\bibinfo{author}{\bibfnamefont{J.~C.} \bibnamefont{Light}},
  \bibinfo{author}{\bibfnamefont{I.~P.} \bibnamefont{Hamilton}},
  \bibnamefont{and} \bibinfo{author}{\bibfnamefont{J.~V.} \bibnamefont{Lill}},
  \bibinfo{journal}{J Chem Phys} \textbf{\bibinfo{volume}{82}},
  \bibinfo{pages}{1400} (\bibinfo{year}{1985}).

\bibitem[{\citenamefont{Corey and Lemoine}(1992)}]{cor92:4115}
\bibinfo{author}{\bibfnamefont{G.~C.} \bibnamefont{Corey}} \bibnamefont{and}
  \bibinfo{author}{\bibfnamefont{D.}~\bibnamefont{Lemoine}},
  \bibinfo{journal}{J. Comp. Phys.} \textbf{\bibinfo{volume}{97}},
  \bibinfo{pages}{4115} (\bibinfo{year}{1992}).

\bibitem[{\citenamefont{Rescigno and McCurdy}(2000)}]{femdvr}
\bibinfo{author}{\bibfnamefont{T.~N.} \bibnamefont{Rescigno}} \bibnamefont{and}
  \bibinfo{author}{\bibfnamefont{C.~W.} \bibnamefont{McCurdy}},
  \bibinfo{journal}{Phys. Rev. A} \textbf{\bibinfo{volume}{62}},
  \bibinfo{pages}{032706} (\bibinfo{year}{2000}).

\bibitem[{\citenamefont{Tolstikhin et~al.}(1996)\citenamefont{Tolstikhin,
  Watanabe, and Matsuzawa}}]{svd}
\bibinfo{author}{\bibfnamefont{O.~I.} \bibnamefont{Tolstikhin}},
  \bibinfo{author}{\bibfnamefont{S.}~\bibnamefont{Watanabe}}, \bibnamefont{and}
  \bibinfo{author}{\bibfnamefont{M.}~\bibnamefont{Matsuzawa}},
  \bibinfo{journal}{J. Phys. B} \textbf{\bibinfo{volume}{29}},
  \bibinfo{pages}{L389} (\bibinfo{year}{1996}).

\end{thebibliography}

\end{document}